\documentclass[english,showpacs,superscriptaddress]{revtex4}

\usepackage[T1]{fontenc}
\usepackage[latin1]{inputenc}
\usepackage{textcomp}
\usepackage{amsmath}
\usepackage{graphicx}
\usepackage{amssymb}
\usepackage{textcomp}

\date{\today}
\makeatletter

\usepackage[usenames,dvipsnames]{color}
\usepackage{bm}
\usepackage[normalem]{ulem}

\usepackage{epsfig}
\usepackage{times}

\newcommand{\GeV}{{\rm GeV}}
\usepackage[hyperref]{hyperref}

\hypersetup{pdfauthor={K. Goeke, V. Guzey, M. Siddikov}, pdftitle={GPDs and DVCS in Color Glass Condensate model}, pdfsubject={High Energy Physics}, pdfkeywords={Nuclear DVCS, Color Glass Condensate, saturation}}

\usepackage{babel}
\makeatother

\begin{document}

\title{Generalized parton distributions and  Deeply Virtual Compton
Scattering in Color Glass Condensate model}

\author{K. Goeke}

\email{Klaus.Goeke@tp2.rub.de}

\affiliation{Institut f\"ur Theoretische Physik II, Ruhr-Universit\"at-Bochum,
D-44780 Bochum, Germany}

\author{V. Guzey}

\email{vguzey@jlab.org}

\affiliation{Theory Center, Jefferson Lab, Newport News, VA 23606, USA}

\author{M. Siddikov}

\email{Marat.Siddikov@tp2.rub.de}

\affiliation{Institut f\"ur Theoretische Physik II, Ruhr-Universit\"at-Bochum,
D-44780 Bochum, Germany}

\affiliation{Theoretical Physics Department, Uzbekistan National University, Tashkent
700174, Uzbekistan}
\preprint{JLAB-THY-08-816}


\begin{abstract}
Within the framework of the Color Glass Condensate model, we evaluate
quark and gluon Generalized Parton Distributions (GPDs)  and the cross section of Deeply Virtual Compton
Scattering (DVCS) in the small-$x_{B}$ region. We demonstrate that
the DVCS cross section becomes independent of energy in the limit
of very small $x_{B}$, which clearly indicates saturation of the
DVCS cross section. Our predictions for the GPDs and the DVCS cross
section at high-energies can be tested at the future Electron-Ion
Collider and in ultra-peripheral nucleus-nucleus
collisions at the LHC. 
\end{abstract}
\maketitle

\section{Introduction}

\label{Sect:CGC}During the last decade hard exclusive reactions,
such as Deeply Virtual Compton Scattering (DVCS), $\gamma^{\ast}(q)+p\to\gamma(q')+p'$,
have been a subject of intensive theoretical and experimental studies~\cite{Mueller:1998fv,Ji:1996nm,Ji:1998pc,Radyushkin:1996nd,Radyushkin:1997ki,Radyushkin:2000uy,Ji:1998xh,Collins:1998be,Collins:1996fb,Brodsky:1994kf,Goeke:2001tz,Diehl:2000xz,Belitsky:2001ns,Diehl:2003ny,Belitsky:2005qn}.
A particular interest has been attached to the generalized
Bjorken kinematics,

\begin{eqnarray}
 &  & -q^{2}=Q^{2}\quad\ {\rm large}\,,\nonumber \\
 &  & W^{2}=(P+q)^{2}\quad\ {\rm large}\,,\nonumber \\
 &  & x_{B}=\frac{Q^{2}}{2\, P\cdot q}={\rm const}\,,\nonumber\\
 &  & t=\Delta^{2}=(P'-P)^{2}\ll Q^{2}\,, \label{eq:Bjorken}\end{eqnarray}
 where $q$ is the momentum of the virtual photon; $P$ is the initial
momentum of the target hadron; $P'$ is the final momentum of
the target, and $t$ is the momentum transfer.

In this kinematics the DVCS amplitude is factorized~\cite{Ji:1998xh,Collins:1998be}
into the convolution of the perturbative coefficient function with
nonperturbative Generalized Parton Distributions (GPDs) of the target.
Recently, the leading-twist dominance (validity of the collinear QCD
factorization) in DVCS on the proton target was demonstrated by the
Hall A collaboration at Jefferson Laboratory~\cite{MunozCamacho:2006hx},
already at rather low values of $Q^{2}$, $1.5\leq Q^{2}\leq2.3$
GeV$^{2}$.

However it turns out that in experiments with nuclei the virtuality
$Q^{2}$ is not always very large, and one cannot say how accurate
 the predictions based on factorization are, or, in other words, how
large  the higher-twist corrections are. One of the examples, where
this approach cannot be applied, is DVCS on the nuclei measured by HERMES collaboration in
DESY~\cite{Ellinghaus:2007dw}. Due to small-$Q^{2}\sim1-2\,\GeV^{2}$ one has to use other effective
models, e.g. Generalized Vector Meson Dominance model (GVMD)~\cite{Goeke:2008rn}.

 At very large $W^{2}$ (very small values
of $x_{B}$), the perturbative collinear factorization is expected
to break down due to high densities of the partons~\cite{Gribov:1984tu}.
Even for relatively large values of $Q^{2}$ when the running coupling
constant of the strong interactions $\alpha_{s}(Q^{2})$ is small,
the effective expansion parameter $\alpha_{s}(Q^{2})g(x,Q^{2})$,
where $g(x,Q^{2})$ is the gluon density in the target, becomes large.
This invalidates the perturbative expansion leading to the collinear
factorization. Since in heavy nuclei the parton densities are enhanced
by the atomic number $A$ compared to those in the nucleon, the onset
of the effects associated with high parton densities may take place
at the values of $x_{B}$, which will be already achieved at the future
Electron-Ion Collider (EIC).

In this paper we use the framework of the Color
Glass Condensate (CGC) model offered in~\cite{McLerran:1993ni,McLerran:1993ka}
(see also recent reviews~\cite{GolecBiernat:2004xu,Iancu:2002aq,Iancu:2003xm,Miller:2001tg}).
We generalize the formalism of the CGC model to exclusive reactions
and evaluate Generalized Parton Distributions (GPDs) and the DVCS
amplitude at small-$x_{B}$. We find that for DVCS off heavy nuclei,
the DVCS cross section is virtually $x_{B}$-independent, i.e.~the
DVCS cross section \textit{saturates} in the small-$x_{B}$ limit. The general saturation property built-in into this model is an essentially nonperturbative effect, which complies with the general Froissart (unitarity) bound~\cite{Froissart:1961ux,Gribov:1968gs,Frankfurt:2001nt}
\begin{eqnarray}
F_{2}(x,Q^{2})\le\ln^{n}\left(s\right)\sim\ln^{n}\left(\frac{1}{x}\right),\label{Froissart:DIS}\end{eqnarray}
where $n=3$ for DIS on nucleons; $n=1$ for DIS on heavy nuclear targets.
This should be compared to the Froissart bound for the case of
hadron-hadron scattering, $\sigma \leq \ln^2 s$.


For comparison, estimates of the $x$-dependence based on the perturbative evolution equations do not possess saturation: The DGLAP predicts a fast growing $x$-dependence~\cite{Bartels:1995iu,Bartels:1996wc}
 \begin{equation}
x\, g(x,\, Q^{2})\sim\exp\left(4\sqrt{\frac{3}{\pi}\alpha_{s}(Q^{2})\ln\frac{Q^{2}}{Q_{0}^{2}}\ln\frac{1}{x}}\right),\label{eq:DGLAP:integrated}\end{equation}
 and the BFKL framework~\cite{DeRujula:1974rf,Kuraev:1976ge,Balitsky:1978ic,Kuraev:1977fs,GolecBiernat:2004xu}
predicts the power-growing $x$-dependence \begin{eqnarray}
x\, g(x,Q^{2})\sim\left(\frac{1}{x}\right)^{4\alpha_{s}(Q^{2})\ln2}\,.\label{eq:BFKL:integrated}\end{eqnarray}

The crucial parameter of the CGC model is the saturation scale $Q_{s}^{2}(x,A)$,
which  gives the threshold for transition to saturation regime. 
The saturation scale $Q_{s}^{2}(x,A)$ comes into play as a universal parameter in many tasks. For example, in CGC explanation of the geometric scaling~\cite{GolecBiernat:1998js} in DIS data from HERA, the structure function $F_{2}(x,Q^{2})$ is represented as a function
of only one variable, \textit{i.e.} $F_{2}(x,Q^{2})=f\left(\frac{Q^{2}}{Q_{s}^{2}(x)}\right)$.

The paper is organized as follows. In Section~\ref{sec:Overview}
we give a brief overview of the model used for evaluations. In particular,
we generalize the original framework of~\cite{McLerran:1993ka,McLerran:1993ni}
to the finite nucleus case in order to consider the off-forward matrix
elements. In Section~\ref{sec:Unintegrated-quark-GPDs} we evaluate
the quark GPDs, and in Section~\ref{sec:DVCS-amplitude} we evaluate
the DVCS amplitude. In Section \ref{sec:Results-and-Discussion} we
present our results and draw conclusions.

\section{Generalized Parton Distributions in the Color Glass Condensate model}

\subsection{Overview of the Color Glass Condensate model}

\label{sec:Overview} The basic assumption of CGC is that one can
separate the partons into fast ($x_{B}\sim1$) and slow ($x_{B}\ll1$)
ones, according to their light-cone fraction $p^{+}$. The former
are considered as classical ``sources'', and the latter are the
dynamical degrees of freedom in the model. In the leading order over
$\alpha_{s}(Q^{2})$ one has just ordinary Yang-Mills equations for
the gluon fields, in NLO one has a standard loop expansion. It is
assumed that dynamics of the ``fast'' partons does not depend
on the ``slow'' partons; thus the configurations of the fast partons
are \textit{random} and one must \textit{average} over all possible
configurations of these ``sources'' $J_{\mu}^{a}(x)=\delta_{\mu+}\rho^{a}(x)$,
where $a$ is a color index, and $x$ is a coordinate. The
weight functional $W[\rho]$ encodes the dynamics of the ``fast''
subsystem and comes as an external parameter in the model. There are
no restrictions on this functional except for the obvious gauge and
Lorentz invariance. An additional requirement of color neutrality,
\begin{equation}
\int d^{3}x\left\langle \rho^{a}(\vec{x})\rho^{b}(\vec{0})\right\rangle =0\,,\label{CGC:ColorNeutrality}\end{equation}

was introduced in~\cite{Lam:1999wu}. It reflects the fact that
the physical states are colorless.

If we define $x_{0}$ as a scale which separates ``fast'' and
``slow'' partons, then the dependence of the functional $W[\rho]$
on the scale $x_{0}$ will be described by a kind of ``renormgroup
equation'' \begin{eqnarray}
\frac{\partial W[\rho;\tau]}{\partial\tau}=\frac{1}{2}\int d\vec{x}d\vec{y}\frac{\delta}{\delta\rho^{a}(\vec{x})}\chi\left(\vec{x},\vec{y}\right)\frac{\delta}{\delta\rho^{b}(\vec{y})}W[\rho;\tau]\,,\end{eqnarray}

where $\tau=\ln\left(\frac{1}{x_{0}}\right)$ and $\chi\left(\vec{x},\vec{y}\right)$
is a complicated functional of the field $\rho$.

While in the general case this equation has not been solved so far, there
are known solutions for some special (asymptotic) cases. Conventionally
$W[\rho]$ is chosen in a Gaussian form~\cite{McLerran:1993ni,McLerran:1993ka,Iancu:2002aq}
\begin{eqnarray}
W[\rho]={\cal N}\exp\left(-\frac{1}{2}\int d\vec{x}d\vec{y}\frac{\rho^{a}(\vec{x})\rho^{a}(\vec{y})}{\lambda(\vec{x},\vec{y})}\right),\label{W:Gaussian:General}\end{eqnarray}
 where ${\cal N}$ is the normalization factor fixed from the condition
$\int{\cal D}\rho W[\rho]=1$ and the function $\lambda(\vec{x},\vec{y})$
is either a constant or a function fixed with some additional assumptions.
Physically the function $\lambda(\vec{x},\vec{y})$ describes correlation
of  partons inside the target. It is obvious that in the infinite
nuclear matter it may depend only on the relative distance, i.e.

\begin{equation}
\lambda(\vec{x},\vec{y})=\mu_{A}^{2}(\vec{x}-\vec{y})\,.\label{eq:MuDefinition}\end{equation}

In the general case, the shape of the function $\mu_{A}^{2}(\vec{r})$
is unknown. However, the color neutrality condition~(\ref{CGC:ColorNeutrality})
and the requirement that in the low parton density limit the model
should reproduce BFKL predictions (\ref{eq:BFKL:integrated}), fix
the short-distance and large-distance behaviour. It was proposed in~\cite{Iancu:2002aq}
that one can use the interpolation

\begin{equation}
\text{Parameterization I:}\qquad\mu_{A}^{2}(\vec{r})=\int\frac{d^{2}k}{(2\pi)^{2}}\mu^{2}(k)e^{-i\vec{k}\vec{r}}=\int\frac{d^{2}k}{(2\pi)^{2}}e^{-i\vec{k}\vec{r}}\frac{k_{\perp}^{2}}{\pi}\frac{\left(\frac{Q_{s}^{2}(x)}{k_{\perp}^{2}}\right)^{\gamma}}{1+\left(\frac{Q_{s}^{2}(x)}{k_{\perp}^{2}}\right)^{\gamma}}\,,\label{CGC:Weight:finitewidth}\end{equation}

where $\gamma=\frac{1}{2}\sqrt{1+\frac{8\ln2}{7\zeta(3)}}\approx0.644$
is a numerical coefficient.

There are also simpler versions of the model~\cite{McLerran:1993ka,McLerran:1993ni},
which neglect correlation of the partons, i.e.

\begin{equation}
\text{Parameterization II:}\qquad\frac{1}{\lambda(\vec{x},\vec{y})}=\frac{\delta(\vec{x}-\vec{y})}{\lambda_{A}(x^{-})}\,,\label{eq:simple-lambdaDefinition}\end{equation}

where $\lambda_{A}(x^{-})$ is some function~\footnote{From physical point of view, the nucleus moving with ultrarelativistic
velocity in laboratory frame is strongly squeezed due to Lorentz contraction,
the observer sees only a thin ``pancake'' with (almost) uniform
width and (almost infinite) radius $R$. This explains the choice
of parameterization~(\ref{eq:simple-lambdaDefinition}), where $\lambda_{A}(x^{-})$
in previous formula must be strongly peaked around $x^{-}\approx0$.
Quite often for simplicity it is assumed that the width might be completely
neglected. A very interesting generalization of the framework to the
finite width case may be found in~\cite{Kovchegov:1996ty}.}. In subsequent sections we will consider first evaluation with a
simple parameterization~(\ref{eq:simple-lambdaDefinition}), and
after that discuss, how the results change for the parameterization
(\ref{CGC:Weight:finitewidth}).

The choice of the Gaussian parameterization~(\ref{W:Gaussian:General})
enables us to evaluate all the results analytically. Notice however,
that~(\ref{W:Gaussian:General}) is explicitly $C$-even, i.e. the
number of quarks is equal to the number of antiquarks \emph{inside
any target} in this model. This  agrees with experimental
fact that the quark and anti-quark parton densities are approximately
equal at small $x_{B}$. On the other hand, $C$-parity of~(\ref{W:Gaussian:General})
implies that the model does not distinguish matter and antimatter
and is not applicable to evaluation of some quantities. For example,
the baryon number and electric charge of the target are exactly zero,
since they are due to the valence quarks. 

An interesting generalization of the Gaussian parameterization~(\ref{W:Gaussian:General}) was discussed in~\cite{Jeon:2005cf}. In particular, it was found that for the model of $k\gg1$ independent noninteracting quarks the distribution is indeed Gaussian, and the first correction is proportional to $\sim d_{abc}\int d^3 x\,\rho^a(\vec x)\rho^b(\vec x)\rho^c(\vec x)$, where $d_{abc}$ is defined from the anticommutator of the generators $T^a$ of the group, $\{T^a, T^b\}= 2d_{abc}T^c$. However, for the DVCS and singlet GPDs discussed in this paper the $C$-odd correction does not contribute.

It is well-known that at high-energies, the real part of scattering
amplitudes is suppressed by the slow energy dependence of the amplitude
compared to the imaginary part~\cite{Gribov:1968uy,Bronzan:1974jh,Sidhu:1974cq}.
Therefore, it is sufficient to consider only the imaginary part. Actually,
as we shall show in Sect.~\ref{sec:DVCS-amplitude}, the real part
of the DVCS amplitude in the CGC model is exactly zero.

The generating functional of the model has a form~\footnote{For the sake of brevity we included only gluons. Addition of the quark
sector is trivial.}

\begin{eqnarray}
Z[j]=\int D\rho\, W[\rho]\frac{\int DA\delta(A^{+})e^{iS[A,\rho]-\int dx j\cdot A}}{\int DA\delta(A^{+})e^{iS[A,\rho]}}\,,\label{CGC:Z}
\end{eqnarray}
where $S[A,\rho]=S[A]+\int d\vec{x}\rho^{a}(\vec{x})A_{-}^{a}(\vec{x})$
and we used light-cone gauge $n\cdot A=0,\, n^{2}=0$. In order to
restore the explicit gauge invariance of the action $S[A,\rho]$, the
interaction term $\int d\vec{x}\rho^{a}(\vec{x})A_{-}(\vec{x})$ is
sometimes replaced with $Tr\int d^{3}\vec{x}\rho(\vec{x})W[A,\vec{x}]$,
where 
\begin{equation}
W[A,\vec{x}]=P\exp\left(ig\int_{-\infty}^{x^{+}}d\zeta A^{+}(\zeta)\right)\label{CGC:Wilson}
\end{equation}
is the Wilson link.

\subsection{Finite nucleus}
\label{sec:FiniteNucleus} Since in this paper we are interested in
DVCS--off-forward reaction, we can no longer use the infinite nuclear
matter approximation. Indeed, the DVCS cross section off a nuclear
target rapidly decreases as one increases the momentum transfer $t$.
As a result, the sizable cross-sections exist only for $|t|\sim1/R_{A}^{2}$,
where $R_{A}$ is the nuclear radius. In the infinite nuclear
matter, all the off-forward cross-sections vanish~\footnote{Technically, we get the prefactors $\delta(\vec{\Delta})$ for all
observables in the original framework of~\cite{McLerran:1993ka,McLerran:1993ni}
in the infinite nuclear matter limit.}. This means that we have to take into account the off-forward kinematics
from the very beginning. If the coordinate of the nucleus center of
mass is $\vec{X}$, then the weight functional $W[\rho]$ may be chosen
as \begin{eqnarray}
 &  & W_{\rho}[\rho,X]=\exp\left\{ -\frac{1}{2}\int d^{3}x\theta(|\vec{x}_{\perp}-\vec{X}_{\perp}|<R_{A})\frac{\rho^{a}(\vec{x}-\vec{X})\rho^{a}(\vec{x}-\vec{X})}{\lambda_{A}(x^{-}-X^{-})}\right\} \,,\label{CGCF:Wrho}\end{eqnarray}
 where we extracted the ``zero mode'' (integration over the nucleus
center of mass) explicitly according to standard technique~\cite{Faddeev:1977rm}
and introduced an explicit cutoff factor $\theta(|\vec{x}_{\perp}-\vec{X}_{\perp}|<R_{A})$
which forbids the \textit{color condensate} $\rho^{a}(\vec{x})$ from
outside of the nucleus. The cutoff in $x^{-}$ is provided
by the factor $\lambda_{A}(x^{-}-X^{-})$. The interaction of gluons
with the condensate is also modified by this cutoff factor:
\begin{equation}
S[A,\rho]=S[A]+Tr\int d^{3}x\theta(|\vec{x}_{\perp}-\vec{X}_{\perp}|<R_{A})\rho(\vec{x})A_{-}(\vec{x})\end{equation}
 for the linear interaction, or \begin{eqnarray}
 &  & S[A,\rho]=S[A]+Tr\int d^{3}x\theta(|\vec{x}_{\perp}|<R_{A})\rho(\vec{x})W[A](\vec{x})\end{eqnarray}
 for the interaction via Wilson link~(\ref{CGC:Wilson}). The generating
functional~(\ref{CGC:Z}) takes the form

\begin{equation}
Z[j]=\int D\rho dX\, e^{i\vec{\Delta}\vec{X}}W[\rho,X]\frac{\int DA\delta(A^{+})e^{iS[A,\rho]-\int dx j\cdot A}}{\int DA\delta(A^{+})e^{iS[A,\rho]}}\,.\end{equation}

Notice that the formal introduction of the $\theta$-functions is
equivalent to the redefinition of the functional integral: \begin{eqnarray}
\int D\rho:=\prod_{x}d\rho(x^{-},|\vec{x}_{\perp}-\vec{X}_{\perp}|<R_{A})d^{3}X\,.\end{eqnarray}
 Indeed, configurations with $\rho(|\vec{x}_{\perp}|>R_{A})\not=0$
do not interact with anything and thus contribute only to the normalization
constant.

Since the coupling constant $\alpha_{s}$ is small, we can take the
integral over the gluon field $A_{\mu}$ in~(\ref{CGC:Z}) in the
saddle-point approximation. In the leading order, the gluon field
$A_{\mu}$ is just the solution of the equation of motion

\begin{eqnarray}
 &  & D_{\mu}F_{a}^{\nu\mu}(x)=\delta^{\nu,+}\delta(x^{-})\rho^{a}(x_{\perp}),\label{eq:EqnMotion}\end{eqnarray}

where $\rho^{a}(x_{\perp})$ is the arbitrary external field, and
an additional gauge constraint $A^{+}=0$ is implied. Notice that
we do not impose any conditions onto the gluonic fields $A_{\mu}^{a}$
at the large distance $|\vec{x}|>R_{A}.$ The solution of the equation~(\ref{eq:EqnMotion})
is~\cite{Iancu:2002aq}

\begin{eqnarray}
 &  & A^{\mu}=U\left(\tilde{A}_{\mu}+\frac{i}{g}\partial_{\mu}\right)U^{\dagger},\label{CGC:General:infinite}\end{eqnarray}
 where~\footnote{The most interesting properties of the solution are the following:
(1) Only transverse components of $A_{\mu}$ are not zero: $A_{k}=\frac{i}{g}U\partial_{k}U^{\dagger},\,\, k=1,2,$.
(2) $A_{i}\to0$ for $x^{-}\to-\infty,$. (3) The only nonzero components
of $F^{\mu\nu}$ are $F^{+k}=-U\partial_{k}\alpha_{a}T^{a}U^{\dagger},$.
(4) In the linear approximation, one should just drop the unitary
matrices $U,U^{\dagger}$ to get $F^{+k}\approx-\partial_{k}\alpha_{a}T^{a}.$ } \begin{eqnarray}
 &  & \tilde{A}_{\mu}=\delta^{\mu+}\alpha(x^{-},x_{\perp})\,,\\
 &  & U=P\exp\left\{ ig\int_{-\infty}^{x^{-}}dz^{-}\alpha_{a}(z^{-},x_{\perp})T^{a}\right\} \,,\\
 &  & \alpha(x^{-},x_{\perp})=\frac{1}{-\partial_{\perp}^{2}}\tilde{\rho}=\int d^{2}y_{\perp}\frac{1}{4\pi}\ln\frac{1}{(x_{\perp}-y_{\perp})^{2}\Lambda_{QCD}^{2}}\tilde{\rho}(x^{-},y_{\perp})\,,\\
 &  & \tilde{\rho}(x^{-},x_{\perp})=U^{\dagger}(x^{-},x_{\perp})\rho(x^{-},x_{\perp})U(x^{-},x_{\perp})\,,\end{eqnarray}
 and $T^{a}$ are the generators of the color group.

The analytical solution~(\ref{CGC:General:infinite}) enables us to evaluate
different correlators. A straightforward evaluation of the $\left\langle \rho\rho\right\rangle $-correlator
with the weight function (\ref{CGCF:Wrho}) yields

\begin{equation}
\langle P'\left|\rho(\vec{x})\rho(\vec{y})\right|P\rangle=\bar{P}^{+}\int d^{3}X\, e^{i\vec{\Delta}\vec{X}}\theta(|\vec{x}_{\perp}-\vec{X}_{\perp}|<R_{A})\lambda_{A}(x^{-}-X^{-})\delta^{3}(\vec{x}-\vec{y})=f(\Delta)\bar{P}^{+}e^{-i\vec{\Delta}\vec{x}}\delta^{3}(\vec{x}-\vec{y})\,,\label{CGCF:rho}\end{equation}
 where \begin{equation}
f(\Delta)=\left(\tilde{\lambda}(\Delta^{+})\equiv\int dx^{-}\lambda(x^{-})e^{-ix^{-}\Delta^{+}}\right)f_{\perp}(\Delta_{\perp})\pi R_{A}^{2},\end{equation}
 and \begin{equation}
f_{\perp}(\Delta_{\perp})=\frac{1}{\pi R_{A}^{2}}\int d^{2}x_{\perp}\theta(|\vec{x}_{\perp}|<R_{A})e^{i\vec{\Delta}_{\perp}\vec{x}_{\perp}}=\frac{J_{1}(\Delta_{\perp}R_{A})}{\Delta_{\perp}R_{A}}.\end{equation}

We can see that for any fixed nonzero $\Delta_{\perp}\not=0$ the
result vanishes in the $R_{A}\to\infty$ limit in agreement with discussion
at the beginning of this section. The evaluation of the gluonic GPDs
defined as~\cite{Goeke:2001tz} \begin{eqnarray}
x\, H^{g}(x,\xi,t)=\frac{1}{\bar{P}^{+}}\int dz^{-}e^{ix\bar{P}^{+}}\left\langle P'\left|F_{+k}^{a}\left(-\frac{z^{-}}{2}\right)F_{+}^{k,a}\left(-\frac{z^{-}}{2}\right)\right|P\right\rangle \end{eqnarray}
is done in quasiclassical approximation, \begin{eqnarray}
x\, H^{g}(x,\xi,t)\approx\int d^{3}X\, e^{i\vec{\Delta}\vec{X}}\int dz^{-}e^{ix\bar{P}^{+}z^{-}}F_{+k}^{a}\left(-\frac{z^{-}}{2}-\vec{X}\right)F_{+}^{k,a}\left(\frac{z^{-}}{2}-\vec{X}\right)\,,\label{CGC:GPDDef}\end{eqnarray}
 where $F_{\mu\nu}$ in the rhs of (\ref{CGC:GPDDef}) corresponds
to the classical solution found in previous subsection. Evaluation
of~(\ref{CGC:GPDDef})~\cite{Iancu:2002aq} gives~\footnote{We omitted here tedious, although quite simple evaluation. An interested
reader may find the logics of evaluation in~\cite{Iancu:2002aq},
generalization to the off-forward case is done in Appendix~\ref{sec:A:GluonGPD} }

\begin{eqnarray}
x\, H^{g}(x,\xi,t) & = & \frac{(N_{c}^{2}-1)}{\bar{P}^{+}}\int d^{3}X\, e^{i\vec{\Delta}\vec{X}}\left(\int d^{3}\tilde{\Delta}\, e^{-i\tilde{\Delta}\vec{X}}\left(-\partial_{r_{\perp}}^{2}\right)\tilde{\gamma}_{A}(x^{-},\vec{r}_{\perp};\tilde{\Delta})\right)\nonumber\\
 & \times & \exp\left[-g^{2}N_{c}\left(\frac{\tilde{f}\left(\vec{0},\frac{\vec{r}}{2}-\vec{X}\right)+\tilde{f}\left(\vec{0},-\frac{\vec{r}}{2}-\vec{X}\right)}{2}-\tilde{f}\left(\vec{r},-\vec{X}\right)\right)\right]_{r_{\perp}\approx1/Q},\label{eq:GluonGPDFinal} \end{eqnarray}

where \begin{equation}
\tilde{f}(\vec{r}_{1},\vec{r}_{2})=\int\frac{d^{2}\tilde{\Delta}}{(2\pi)^{2}}e^{-i\tilde{\Delta}\vec{r_{2}}}\int_{-\infty}^{+\infty}dz^{-}\tilde{\gamma}_{A}(z^{-},\vec{r}_{1};\tilde{\Delta}),\end{equation}
 and $\tilde{\gamma}_{A}(x^{-},\vec{r}_{\perp})$ is defined as

\begin{equation}
f(\Delta)\bar{P}^{+}\int\frac{d^{3}k}{(2\pi)^{3}}\frac{e^{-ix(k+\Delta/2)}e^{iy(k-\Delta/2)}}{\left(k_{\perp}-\frac{\Delta_{\perp}}{2}\right)^{2}\left(k_{\perp}+\frac{\Delta_{\perp}}{2}\right)^{2}}=\delta(x^{-}-y^{-})\tilde{\gamma}_{A}(x^{-},\vec{x}_{\perp}-\vec{y}_{\perp})e^{i\vec{\Delta}(\vec{x}+\vec{y})/2}\,.\label{CGCF:alpha:linear}\end{equation}
 As one can see from~(\ref{eq:GluonGPDFinal}), the gluon GPD $H^{g}(x,\xi,t)$
has a trivial $x$-dependence $1/x$ for all $(\xi,t)$, since $x$ does not enter the right-hand side of Eq.~(\ref{eq:GluonGPDFinal}). Physically,
the exponent in (\ref{eq:GluonGPDFinal}) takes into account nonlinear
in $\alpha_{s}$ effects in the model.

\subsection{Alternative kernel}

\label{Sect:CGCA} In this section we discuss how all the previous
formulae change with an alternative weight function~(\ref{CGC:Weight:finitewidth}).
The weight functional in this case should be written as \begin{eqnarray}
 &  & W_{\rho}[\rho,\, X]={\cal N}\exp\left\{ -\frac{1}{2}\int d^{3}x\,\theta\left(\left|\vec{x}_{\perp}-\vec{X}_{\perp}\right|<R_{A}\right)\frac{\rho^{a}(\vec{x}-\vec{X})\rho^{a}(\vec{y}-\vec{X})}{\mu_{A}^{2}(\vec{x}-\vec{y})}\right\}\,, \label{CGCA:Wrho}\end{eqnarray}
 where the function $\mu_{A}^{2}(\vec{z})$ describes correlation
of hadrons inside the nuclei and was defined in (\ref{CGC:Weight:finitewidth}).
Performing the evaluation as was discussed in Sect.~\ref{sec:FiniteNucleus},
we obtain 
\begin{eqnarray}
 &  & \langle P'\left|\rho(\vec{x})\rho(\vec{y})\right|P\rangle=\int dXe^{i\vec{\Delta}\vec{X}}\theta\left(\left|\vec{x}_{\perp}-\vec{X}_{\perp}\right|\le R_{A}\right)\theta\left(\left|\vec{y}_{\perp}-\vec{X}_{\perp}\right|\le R_{A}\right)\mu_{A}^{2}(\vec{x}_{\perp}-\vec{y}_{\perp})=\nonumber\\
 &  & =f(\vec{x}-\vec{y},\Delta)e^{i\vec{\Delta}\frac{\vec{x}+\vec{y}}{2}}\mu_{A}^{2}(\vec{x}-\vec{y})=\int\frac{d^{3}k}{(2\pi)^{3}}\tilde{\mu}_{A}^{2}(\vec{k})e^{-ix(k+\Delta/2)}e^{iy(k-\Delta/2)}\,,\label{CGCA:rho}
\end{eqnarray}

where\begin{eqnarray}
\tilde{\mu}_{A}^{2}(\vec{k}) & = & \int d^{2}\rho e^{-i\vec{k}\vec{\rho}}f(\vec{\rho},\Delta)\mu_{A}^{2}(\vec{\rho}),\label{CGCA:mu2}\\
f(\vec{\rho},\Delta) & = & \int d^{2}Xe^{i\vec{\Delta}\vec{X}}\theta\left(\left|\frac{\vec{\rho}}{2}-\vec{X}_{\perp}\right|\le R_{A}\right)\theta\left(\left|\frac{\vec{\rho}}{2}+\vec{X}_{\perp}\right|\le R_{A}\right)=\label{CGCA:f}\\
 & = & \int\frac{d^{2}k}{(2\pi)^{2}}e^{-i\vec{k}\vec{\rho}}\phi\left(\vec{k}+\frac{\vec{\Delta}}{2}\right)\phi\left(\vec{k}-\frac{\vec{\Delta}}{2}\right),\nonumber \end{eqnarray}

\begin{equation}
\phi(\vec{k})=\pi R_{A}^{2}\frac{2J_{1}(kR_{A})}{kR_{A}}.\label{CGCA:Phi}\end{equation}

From (\ref{CGCA:rho}) we can see that in the finite nuclei the color
neutrality condition~(\ref{CGC:ColorNeutrality}) implies that we
have to identify $\mu_{A}^{2}$ from~(\ref{CGC:Weight:finitewidth})
with $\tilde{\mu}_{A}^{2}(r).$ For $\tilde{\gamma}_{A}(x^{-},\vec{r}_{\perp})$
we can immediately obtain \begin{equation}
\langle P'\left|\alpha(\vec{x})\alpha(\vec{y})\right|P\rangle=...=\int\frac{d^{3}k}{(2\pi)^{3}}\tilde{\mu}_{A}^{2}(k)\frac{e^{-ix(k+\Delta/2)}e^{iy(k-\Delta/2)}}{\left(k_{\perp}-\frac{\Delta_{\perp}}{2}\right)^{2}\left(k_{\perp}+\frac{\Delta_{\perp}}{2}\right)^{2}}=\delta(x^{-}-y^{-})\tilde{\gamma}_{A}(x^{-},\vec{x}_{\perp}-\vec{y}_{\perp})e^{i\vec{\Delta}(\vec{x}+\vec{y})/2}\,.\label{CGCA:alpha}\end{equation}
 Thus we can see that this kernel differs from the previous one only
by an additional factor $\tilde{\mu}^{2}(\vec{k})$ in the integrand.

\subsection{Quark propagator in CGC field}

\label{sub:QuarkPropagator}

\begin{figure}[h]
\includegraphics[scale=0.3]{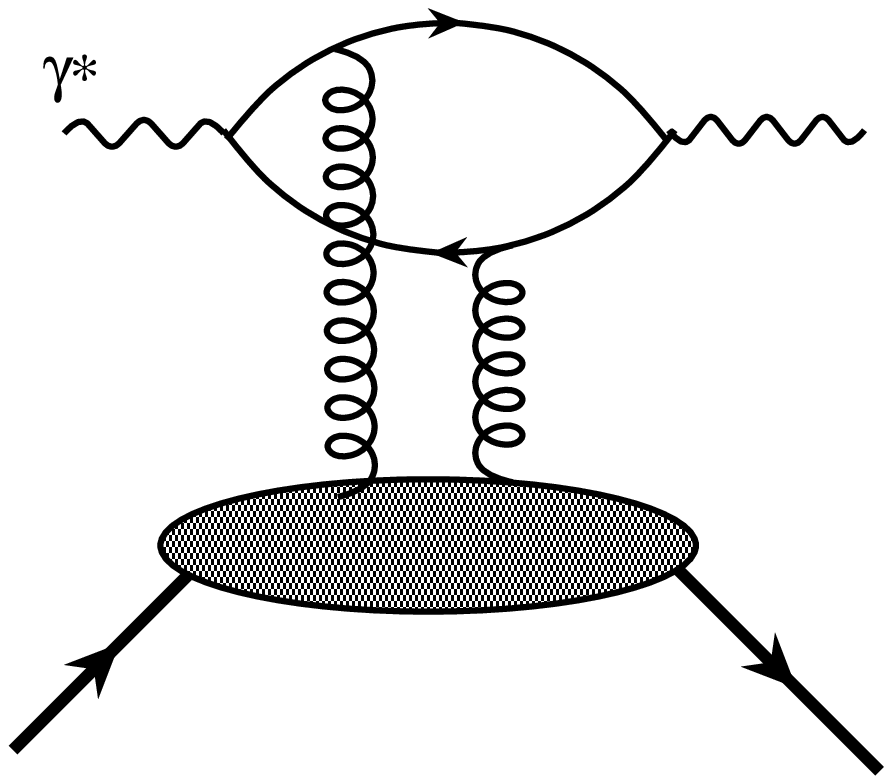} 
\includegraphics[scale=0.3]{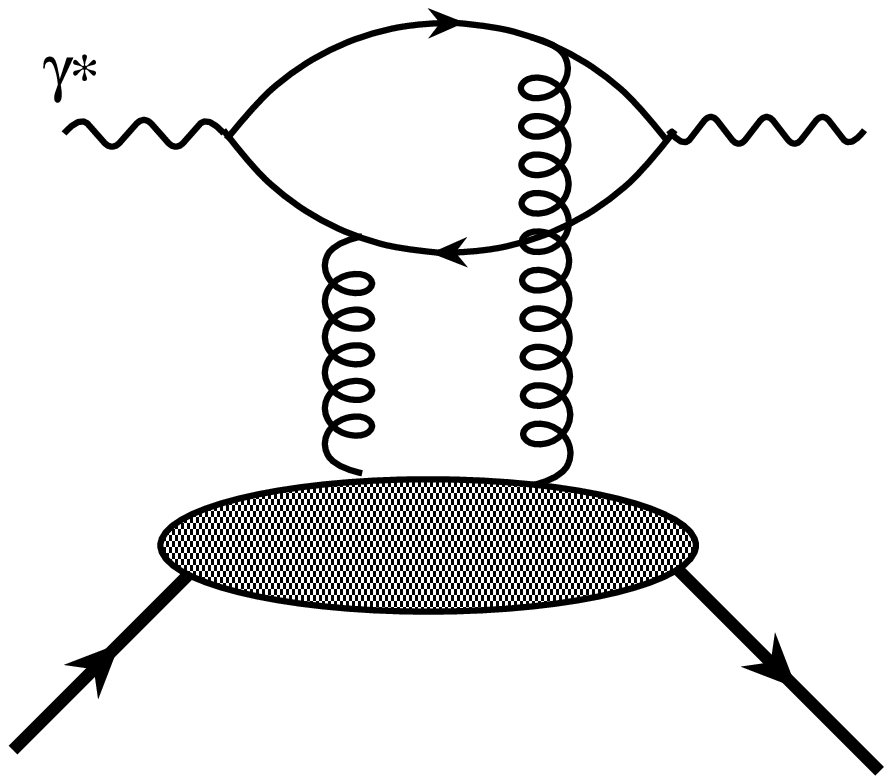}
\includegraphics[scale=0.3]{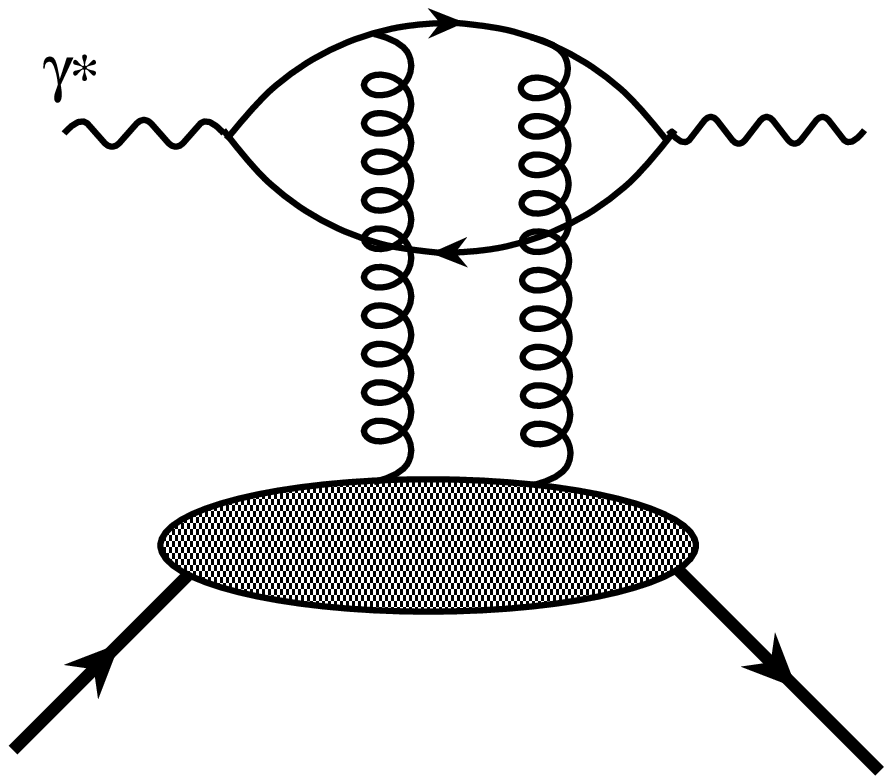} 
\includegraphics[scale=0.3]{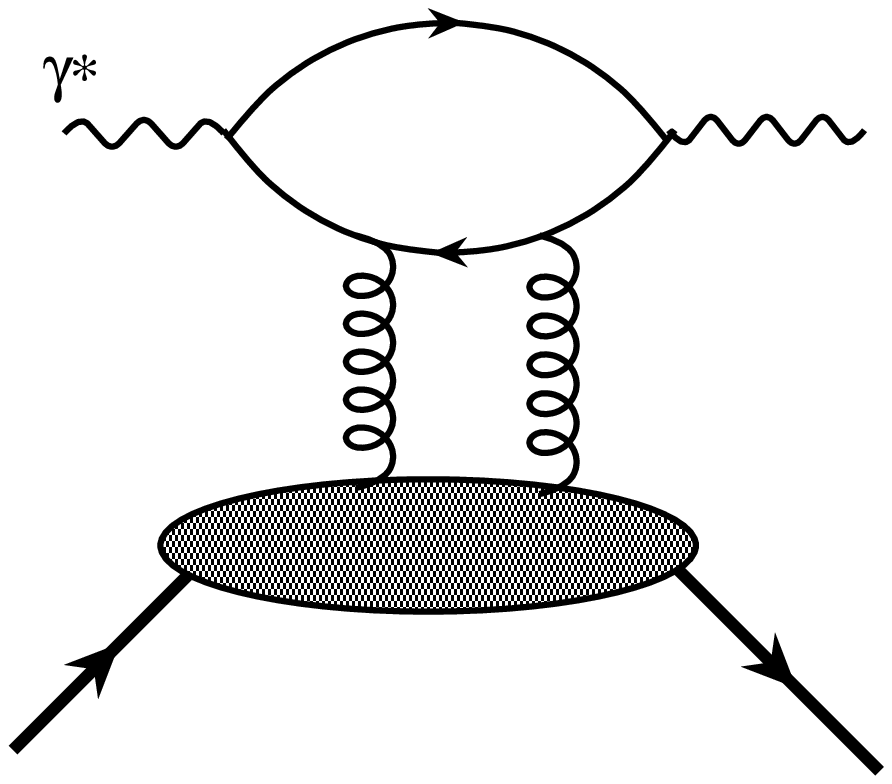}

\caption{\label{CGC:Fig:ColorDipole} Diagrams contributing to DVCS in color
dipole approximation. See~\cite{Kowalski:2003hm} for an example of
DIS evaluation in this approach.}

\end{figure}

Although for the evaluation of the DVCS amplitude one may use the
color dipole approximation, in this paper we evaluate the GPDs and
Compton amplitudes directly. In the diagrammatic language this corresponds
to summation of all~\footnote{To avoid ambiguity, this is ``quenched approximation'', where
the diagrams with gluon loops are not taken into account. Gluon loops
correspond to ${\cal O}(\alpha_{s})$-corrections.} the multigluon diagrams, whereas the color dipole approach assumes
either only Born term contribution or Eikonal approximation, as is
shown on the Fig.~\ref{CGC:Fig:ColorDipole}.

For the evaluation of the quark GPDs in the leading order over $\alpha_{s}(Q^{2}),$
we need to evaluate the quark propagator in the classical gluonic
field found in the previous section. To this end, we consider only
the zero width limit,\begin{equation}
\rho=\delta(x^{-})\rho(\vec{x}_{\perp}).\label{eq:rho:zerowidth}\end{equation}
 Beyond this limit, equations with explicit $x^{-}$-dependence become
much more complicated. Physically, the use of~(\ref{eq:rho:zerowidth})
in the off-forward kinematics is justified, since the light-cone fractions
of the partons are small, i.e. $x,\xi\ll1$.

The basic idea is that for $x^{-}\not=0$ the field $\rho(\vec{x})=0$
and we have just vacuum equations, gluon field $A_{\mu}$ reduces
to a pure gauge. It is possible to choose the gauge in such a way
that for $x^{-}<0$ the field disappears, $A_{\mu}=0$, and for $x^{-}>0$
it is a pure gauge, $A_{\mu}=\frac{i}{g}U\partial_{\mu}U^{\dagger}$
and thus the wave function of the quark has a form 
\begin{equation}
\psi_{ps}(x)=\left\{ 
\begin{array}{l}
u_{s}(p)e^{-ipx},x^{-}<0\\
\\
\int d^{4}p'\delta(p^{2}-p'^{2})\sum_{s'}C_{ss'}(p,p')u_{s'}(p')e^{-ipx},x^{-}>0
\end{array}\right.\,,
\end{equation}
 where $u_{s}(p)$ is a free Dirac spinor, and the matrix $C_{ss'}(p,p')$
is found from the continuity at the point $x^{-}=0$. One subtle point
is that the Dirac operator has the form \begin{equation}
i\hat{D}=i\partial_{-}\gamma^{-}+...\,,\end{equation}

and matrix $\gamma^{-}$ is singular, because it is proportional to
the light-cone projector $\Lambda^{(-)}$. This implies that the continuity
condition must be imposed not on the function $\psi_{ps}(x)$ as a
whole, as in~\cite{McLerran:1993ka,McLerran:1993ni,Iancu:2002aq},
but rather only on the component~\footnote{Thanks to P. Pobylitsa. Attempt to impose the continuity on both components
$\psi_{ps}^{(+)}$ and $\psi_{ps}^{(-)}$ leads to inconsistency with
equations of motion. See~\cite{Pobylitsa:unpublished} for more
details.} $\psi_{ps}^{(-)}(x)=\Lambda^{(-)}\psi_{ps}(x)$. The final result
for the wave function is 
\begin{eqnarray}
&  & \psi_{ps}(x)=\theta(-x^{-})u_{s}(p)e^{-ip\cdot x}\nonumber\\
&  &+ \theta(x^{-})U(x_{\perp})\int\frac{d^{4}k}{(2\pi)^{4}}\delta(k^{-}-p^{-})\delta\left(k^{+}-\frac{k_{\perp}^{2}+p^{2}}{2p^{-}}\right)\nonumber \\
&  &\times \left(\int d^{2}ze^{i(p_{\perp}-k_{\perp})\cdot z}U^{\dagger}(z)\right)e^{-ik\cdot x}\left(1+\frac{\gamma_{0}}{k^{-}\sqrt{2}}(\hat{k}_{\perp}+M)\right)\Lambda^{(-)}u_{s}(p), \label{CGCF:Solution}
\end{eqnarray}
 where $M$ is the mass of the quark. The evaluation of the
quark propagator according to \begin{eqnarray}
S(x,y)=\int\frac{d^{4}p}{(2\pi)^{4}}\frac{\sum_{s}\psi_{ps}(x)\bar{\psi}_{ps}(y)}{p^{2}-M^{2}+i0}\,,\end{eqnarray}
 yields \begin{eqnarray}
 &  & S(x,y)-S_{0}(x-y)=\nonumber\\
 &  & =\left\{ \begin{array}{ll}
0, & x^{-}<0,y^{-}<0\\
\left(U(x_{\perp})U^{\dagger}(y_{\perp})-1\right)S_{0}(x-y), & x^{-}>0,y^{-}>0\\
\int\frac{d^{4}p}{(2\pi)^{4}}\frac{1}{p^{2}-M^{2}+i0}\int\frac{d^{2}k}{(2\pi)^{2}}\exp\left\{ i\left(\frac{k_{\perp}^{2}+M^{2}}{2p^{-}}y^{-}+p^{-}y^{+}-k_{\perp}y_{\perp}-p\cdot x\right)\right\} \times\\
\int d^{2}ze^{-i(p_{\perp}-k_{\perp})z}(\hat{p}+M)\Lambda^{(+)}\left(1+\frac{\gamma_{0}}{p^{-}\sqrt{2}}(M-\hat{k}_{\perp})\right)\left(U(z)U^{\dagger}(y_{\perp})-1\right), & x^{-}<0,y^{-}>0\\
\int\frac{d^{4}p}{(2\pi)^{4}}\frac{1}{p^{2}-M^{2}+i0}\int\frac{d^{2}k}{(2\pi)^{2}}\exp\left\{ -i\left(\frac{k_{\perp}^{2}+M^{2}}{2p^{-}}x^{-}+p^{-}x^{+}-k_{\perp}x_{\perp}-p\cdot y\right)\right\} \times\\
\int d^{2}ze^{i(p_{\perp}-k_{\perp})z}\left(1+\frac{\gamma_{0}}{p^{-}\sqrt{2}}(\hat{k}_{\perp}+M)\right)\Lambda^{(-)}(\hat{p}+M)\left(U(x_{\perp})U^{\dagger}(z)-1\right), & x^{-}>0,y^{-}<0\end{array}\right. \label{CGCF:Propagator:Final} \end{eqnarray}
 where $S_{0}(x-y)$ is the free propagator \begin{equation}
S_{0}(x-y)=\int\frac{d^{4}p}{(2\pi)^{4}}\frac{e^{-ip(x-y)}}{\hat{p}-M+i0}.\end{equation}

It might be checked that the propagator $S(x,y)$ satisfies the equation
$(i\hat{D}-M)S(x,y)=\delta(x-y)$ as well as reduces to $S_{0}(x-y)$
in the $U\to1$ limit.

\section{Unintegrated quark GPDs}

\label{sec:Unintegrated-quark-GPDs}

In this section we evaluate unintegrated quark GPDs defined via the
following matrix element (we assume that the target has spin $0$)
\begin{eqnarray}
 &  & H(x,\xi,\vec{\Delta}_{\perp},\vec{k}_{\perp})=\label{CGC:H:Unintegrated}\\
 &  & =\int\frac{dz^{-}}{2\pi}\int d^{2}r_{\perp}e^{-i\vec{k}_{\perp}\vec{r}_{\perp}}e^{ix\bar{P}^{+}z^{-}}\left\langle P'\left|\bar{\psi}\left(-\frac{z^{-}}{2}-\frac{\vec{r}_{\perp}}{2}\right)\gamma^{+}\psi\left(\frac{z^{-}}{2}+\frac{\vec{r}_{\perp}}{2}\right)\right|P\right\rangle \,.\nonumber \end{eqnarray}
 In the forward limit ($\vec{\Delta}_{\perp}\to0,\xi\to0$) the function
$H\left(x,\xi,\vec{\Delta}_{\perp},\vec{k}_{\perp}\right)$ reduces
to unintegrated parton distribution $q\left(x,\vec{k}_{\perp}\right)$,
and when integrated over $\vec{k}_{\perp}$, it gives ordinary GPDs.
In the quasiclassical approximation, (\ref{CGC:H:Unintegrated}) reduces
to \begin{eqnarray}
 &  & H(x,\xi,\vec{\Delta}_{\perp},\vec{k}_{\perp})=\label{CGC:H:Unintegrated:quasiclassical}\\
 &  & =\int\frac{dz^{-}}{2\pi}e^{ix\bar{P}^{+}z^{-}}\int d^{2}r_{\perp}e^{-i\vec{k}_{\perp}\vec{r}_{\perp}}i\bar{P}^{+}\int d^{3}Xe^{-i\vec{\Delta}\vec{X}}\left\langle Tr\left[\gamma^{+}S\left(-\frac{z^{-}}{2}-\frac{\vec{r}_{\perp}}{2}-\vec{X},\frac{z^{-}}{2}+\frac{\vec{r}_{\perp}}{2}-\vec{X}\right)\right]\right\rangle \,,\nonumber \end{eqnarray}
 where here and below angular brackets without explicit initial
and final states $\left\langle ...\right\rangle $ are the short-hand
notation for averaging (integration) over all possible configurations
$\rho(x)$, \textit{i.e.} $\left\langle \hat{O}\right\rangle :=\int\mathcal{D}\rho\, W[\rho]O(\rho)$.
Substituting the propagator (\ref{CGCF:Propagator:Final}) and taking
the integral over each domain, one obtains the final result \begin{equation}
H\left(x,\xi,\vec{\Delta}_{\perp},\vec{k}_{\perp}\right)=H^{(+-)}+H^{(-+)}\label{CGC:H:Unintegrated:Final}\end{equation}
 where \begin{eqnarray}
H^{+-} & = & 2\, N_{c}\int\frac{d^{2}\kappa^{\perp}}{(2\pi)^{2}}\tilde{\gamma}\left(\kappa^{\perp}+\frac{\Delta^{\perp}}{2},\kappa^{\perp}-\frac{\Delta^{\perp}}{2}\right)\nonumber\\
 &  & \times\frac{M^{2}-\left(\vec{k}+\frac{\vec{\Delta}_{\perp}}{2}\right)\cdot\left(\vec{k}-\vec{\kappa}_{\perp}\right)}{(x-\xi)\left(\left(\vec{k}-\vec{\kappa}_{\perp}\right)^{2}+M^{2}\right)-(x+\xi)\left(\left(\vec{k}+\frac{\vec{\Delta}_{\perp}}{2}\right)^{2}+M^{2}\right)}\ln\left|\frac{x-\xi}{x+\xi}\,\frac{\left(\vec{k}-\vec{\kappa}_{\perp}\right)^{2}+M^{2}}{\left(\vec{k}+\frac{\vec{\Delta}_{\perp}}{2}\right)^{2}+M^{2}}\right|\,,\\
H^{-+} & = & 2\, N_{c}\int\frac{d^{2}\kappa^{\perp}}{(2\pi)^{2}}\tilde{\gamma}\left(\kappa^{\perp}+\frac{\Delta^{\perp}}{2},\kappa^{\perp}-\frac{\Delta^{\perp}}{2}\right)\nonumber\\
 &  & \times\frac{M^{2}-\left(\vec{k}-\frac{\vec{\Delta}}{2}\right)\cdot(\vec{k}_{\perp}-\vec{\kappa}_{\perp})}{(x+\xi)\left(\left(\vec{k}_{\perp}-\vec{\kappa}_{\perp}\right)^{2}+M^{2}\right)-(x-\xi)\left(\left(\vec{k}-\frac{\vec{\Delta}}{2}\right)^{2}+M^{2}\right)}\ln\left|\frac{x+\xi}{x-\xi}\,\frac{\left(\vec{k}_{\perp}-\vec{\kappa}_{\perp}\right)^{2}+M^{2}}{\left(\vec{k}-\frac{\vec{\Delta}}{2}\right)^{2}+M^{2}}\right|\,,\end{eqnarray}
 the superscript signs $(\pm,\pm)$ refer to different integration
domains over $(dX^{-},dz^{-})$ in (\ref{CGC:H:Unintegrated:quasiclassical}),
and function $\tilde{\gamma}\left(\vec{\kappa}-\frac{\vec{\Delta}_{\perp}}{2},\vec{\kappa}+\frac{\vec{\Delta}_{\perp}}{2}\right)$
is defined as \begin{equation}
\tilde{\gamma}\left(\kappa^{\perp}+\frac{\Delta^{\perp}}{2},\kappa^{\perp}-\frac{\Delta^{\perp}}{2}\right):=\int\frac{d^{2}\rho d^{2}X}{(2\pi)^{2}}e^{i\Delta^{\perp}X^{\perp}+i\kappa^{\perp}\rho}\left\langle U^{\dagger}\left(X+\frac{\rho}{2}\right)U\left(X-\frac{\rho}{2}\right)\right\rangle \label{tildeGamma:definition}\end{equation}
 Evaluation of this quantity (see Sect.\ref{sec:A:GluonGPD} for details)
yields \begin{eqnarray}
 &  & \tilde{\gamma}\left(\vec{\kappa}-\frac{\vec{\Delta}_{\perp}}{2},\vec{\kappa}+\frac{\vec{\Delta}_{\perp}}{2}\right)=\\
 &  & \int d^{2}r\, e^{i\vec{\kappa}\vec{r}}\int d^{2}X_{\perp}e^{i\vec{\Delta}_{\perp}\vec{X}_{\perp}}\exp\left[-g^{2}N_{c}\left(\frac{\tilde{f}\left(\vec{0},\frac{\vec{r}}{2}-\vec{X}\right)+\tilde{f}\left(\vec{0},-\frac{\vec{r}}{2}-\vec{X}\right)}{2}-\tilde{f}\left(\vec{r},-\vec{X}\right)\right)\right].\nonumber \end{eqnarray}
 Notice that GPD~(\ref{CGC:H:Unintegrated:Final}) is antisymmetric,
\textit{i.e.} $H(-x,\xi)=-H(-x,\xi)$. Also, (\ref{CGC:H:Unintegrated:Final})
is not required to satisfy polynomiality since the original model
is valid only for $x\ll1$.

One of the subtle points of the result~(\ref{CGC:H:Unintegrated:Final})
is the logarithmic behaviour $\sim\ln|x\pm\xi|$ in the vicinity of
the points $x\sim\pm\xi$. Physically, in these points one of the
quarks has a zero light-cone fraction and becomes especially sensitive
to the details of the model. However, since we are in a saturation
regime, factorization formula does not work and we expect that such
behaviour should not cause any physical problems. In Appendix~\ref{sec:A:QuarkGPD}
we give details of evaluation of~(\ref{CGC:H:Unintegrated:Final}),
and in particular discuss the logarithmic singularities.

\section{DVCS amplitude}

\label{sec:DVCS-amplitude}

In this section we evaluate the DVCS amplitude directly (not using
factorization). The first reason for this is that the GPDs evaluated
in the previous section are valid only for the small $x\ll1$ whereas
the convolution formula which follows from factorization implies integration
over the light-cone fraction over the region $-1<x<1$. The second
reason is that, as we discussed in Sect.~(\ref{Sect:CGC}), in the
saturation (high-density) regime the convolution formula becomes invalid.

The starting point of our derivation is the definition of the DVCS
amplitude \begin{eqnarray}
A_{\mu\nu}=-i\int d^{4}z\left\langle P'\left|J_{\nu}(0)J_{\mu}(z)\right|P\right\rangle _{A}e^{-iq\cdot z}\,.\label{CGCF:DVCS:Definition}\end{eqnarray}
 In the quasiclassical approximation the matrix element $\left\langle P'\left|J_{\nu}(0)J_{\mu}(z)\right|P\right\rangle _{A}$
is reduced to \begin{eqnarray}
 &  & \left\langle P'\left|J_{\nu}(0)J_{\mu}(z)\right|P\right\rangle _{A}=\int d^{3}Xe^{i\vec{\Delta}\vec{X}}\left\langle P'\left|J_{\nu}(-\vec{X})J_{\mu}(z-\vec{X})\right|P\right\rangle \nonumber\\
 &  & =-\int d^{3}Xe^{i\vec{\Delta}\vec{X}}\left\langle Tr\left[\gamma_{\mu}S(z-X,-X)\gamma_{\nu}S(-X,z-X)\right]\right\rangle \label{CGCF:DVCS:Matrix}\end{eqnarray}
 where $S(x,y)$ is the propagator~(\ref{CGCF:Propagator:Final}).
Substituting~(\ref{CGCF:Propagator:Final}) into~(\ref{CGCF:DVCS:Matrix})
and taking the integrals, we may reduce the DVCS amplitude to the
form \begin{eqnarray}
 &  & A_{\mu\nu}=i\frac{M_{A}}{2\pi}\int\frac{d^{3}p}{(2\pi)^{3}}\frac{d^{2}k}{(2\pi)^{2}}\frac{\Theta\left(-\frac{q^{-}}{2}\le p^{-}\le\frac{q^{-}}{2}\right)}{q^{+}\left((p^{-})^{2}-(q^{-})^{2}/4\right)+\frac{k_{\perp}^{2}+M^{2}}{2}q^{-}-i0}\tilde{\gamma}\left(\vec{k}-\frac{\vec{\Delta}_{\perp}}{2},\vec{k}+\frac{\vec{\Delta}_{\perp}}{2}\right)\nonumber\\
 &  &\times \frac{1}{2(q^{+}-\Delta^{+})\left((p^{-})^{2}-(q^{-})^{2}/4\right)+p^{-}\vec{p}_{\perp}\vec{\Delta}_{\perp}+q^{-}\left(\vec{p}_{\perp}^{2}+M^{2}+\frac{\vec{\Delta}_{\perp}^{2}}{4}\right)}\nonumber \\
 &  &\times \left(8N_{c}\delta_{\mu+}\delta_{\nu+}\left(M^{2}-\vec{k}_{\perp}^{2}-\vec{p}_{\perp}^{2}\right)\left(M^{2}-\vec{p}_{\perp}^{2}+\frac{\vec{\Delta}_{\perp}^{2}}{4}\right)\right.\nonumber \\
 &  &+ 8N_{c}\delta_{\mu\perp}\delta_{\nu\perp}\left[p_{\perp}^{\mu}p_{\perp}^{\nu}\left((q^{-})^{2}-4(p^{-})^{2}\right)+p^{-}g_{\mu\nu}\left(4p^{-}(M^{2}+p_{\perp}^{2})-q^{-}\vec{p}_{\perp}\cdot\vec{\Delta}_{\perp}\right)\right]\nonumber \\+
 &  & 32N_{c}\delta_{\mu-}\delta_{\nu-}\left((p^{-})^{2}-\frac{(q^{-})^{2}}{4}\right)^{2}\nonumber \\
 &  &+ \left.8N_{c}(\delta_{\mu+}\delta_{\nu-}+\delta_{\nu+}\delta_{\mu-})\left((p^{-})^{2}-\frac{(q^{-})^{2}}{4}\right)\left(2M^{2}-2\vec{p}_{\perp}^{2}-\vec{k}_{\perp}^{2}+\frac{\vec{\Delta}_{\perp}^{2}}{4}\right)\right)\label{CGCF:DVCS:Final}\,. \end{eqnarray}
 One interesting point is that the real part of~(\ref{CGCF:DVCS:Final})
is \textit{exactly} zero. Indeed, taking the imaginary part of the
first ratio containing $-i0$ and using \begin{eqnarray}
\frac{1}{x-i0}=P\left(\frac{1}{x}\right)+i\pi\delta(x),\end{eqnarray}
 we can immediately find that the argument of $\delta$-function is
zero only for \begin{eqnarray}
|p^{-}|=\frac{q^{-}}{2}\sqrt{1+\frac{2(k_{\perp}^{2}+M^{2})}{Q^{2}}}\ge\frac{q^{-}}{2},\end{eqnarray}
 i.e. outside the integration domain. For comparison, from phenomenology
it is known that the high-energy amplitude gets dominant contribution
from the imaginary part.

\section{Results for GPDs and DVCS cross-sections}

\label{sec:Results-and-Discussion} In this section we present results
of the numerical evaluation of the GPDs and DVCS cross-sections. In
subsection~(\ref{subsec:Model-I}) we consider first the results
with a simpler parameterization~(\ref{eq:simple-lambdaDefinition}),
and after that in subsection~(\ref{subsec:Model-II}) with a more
realistic parameterization~(\ref{CGC:Weight:finitewidth}).

\subsection{Results with parameterization II}

\label{subsec:Model-I}

As one can see from (\ref{eq:GluonGPDFinal}), for both parameterizations
I and II the $x$-dependence of the gluon GPD $H_{A}^{g}(x,\xi,t)$
is trivial--just $1/x$ for all $(\xi,\, t)$. For quark GPDs $H_{A}^{g}(x,\xi,t)$
the $x$-dependence is more complicated, however in the forward case
the parton distribution $q_{A}(x)$ has also a simple $1/x$-dependence.
For better understanding, we prefer to discuss out results for the
gluons in terms of the ratio $H_{A}^{g}(x,\xi,t)/g_{A}(x),$ which
measures the off-forward effects, and $g_{A}(x)$ is the forward
gluon PDF evaluated in the same model.

\begin{figure}[h]
\includegraphics[scale=0.3]{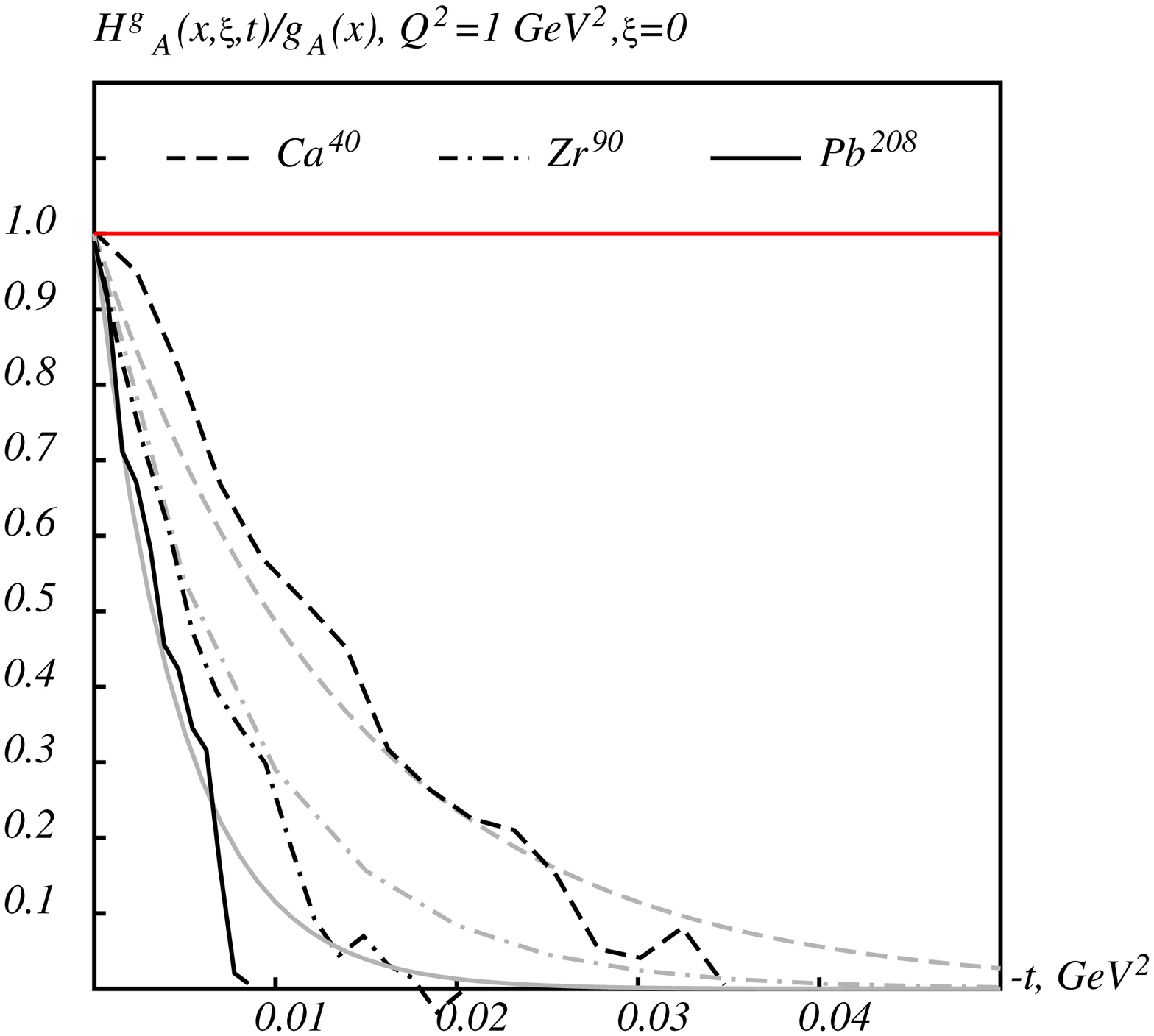} 
\includegraphics[scale=0.3]{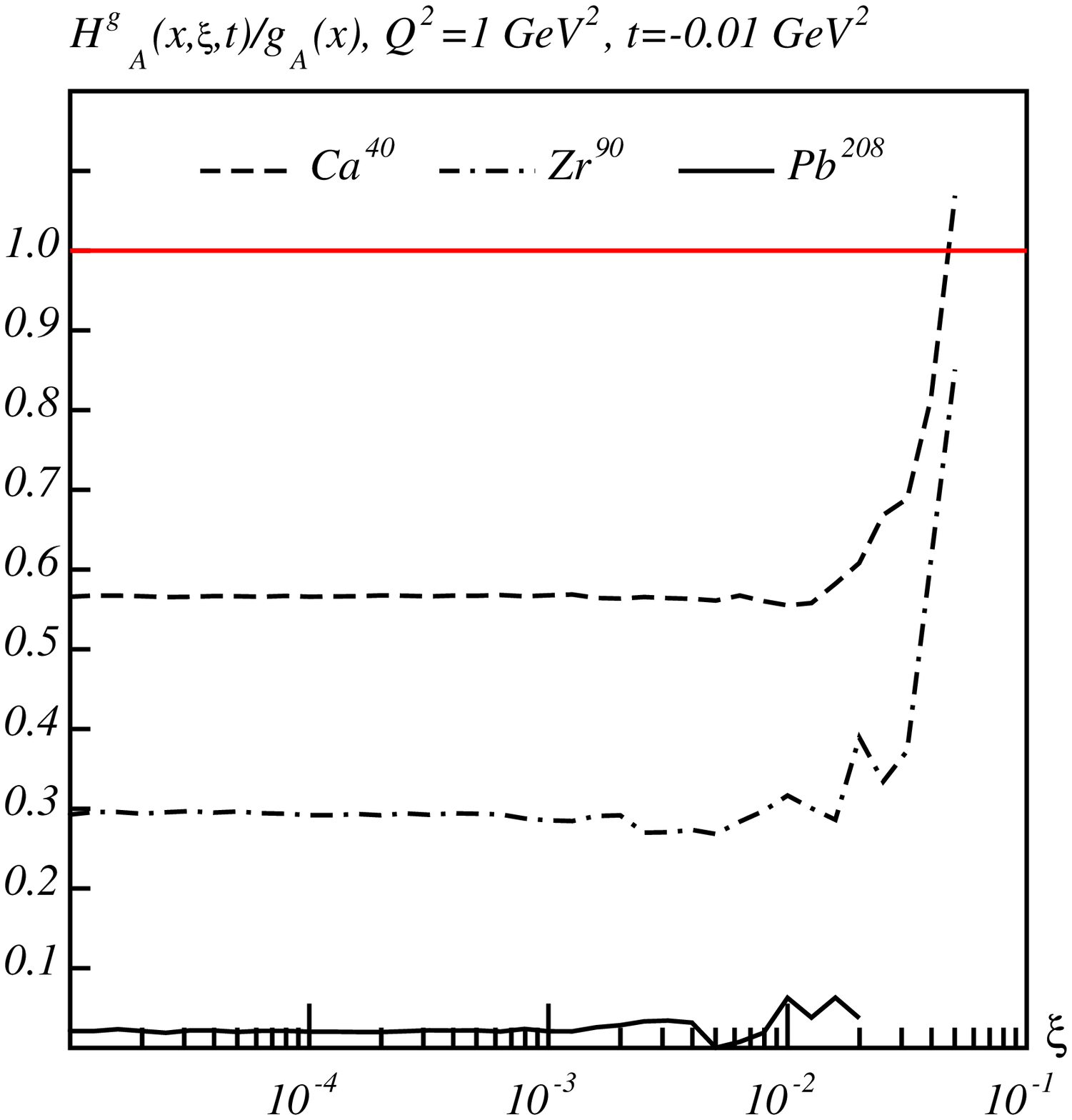}

\caption{\label{CGC:Results:G_dependence} Left plot: $t$-dependence of the
gluon distribution for different nuclei. $\xi=0,Q^{2}=1\,\GeV^{2}$.
For comparison, we also plotted in grey lines the nuclear formfactor
in the simplest exponential parameterization $F_{A}(t)=\exp\left(\frac{R_{A}^{2}}{6}t\right)$.
Right plot: $\xi$-dependence for the same nuclei for fixed $t=-0.01\,\GeV^{2},Q^{2}=1\,\GeV^{2}$.
We do not plot the $x$-dependence of the gluon GPD $H^{g}$, which
is according to~(\ref{eq:GluonGPDFinal}) just a trivial $1/x$ for
all $(\xi,t)$.}

\end{figure}

In Figure~\ref{CGC:Results:G_dependence} we plot the $\xi$ and
$t$-dependence of the ratio $H_{A}^{g}(x,\xi,t)/g_{A}(x)$ for different
nuclei. In the left panel of  Figure~\ref{CGC:Results:G_dependence}
we plot the $t$-dependence of the ratio $H_{A}^{g}(x,\xi,t)/g_{A}(x)$
in nuclei for $\xi=0$. We can see that $H^{g}(x,\xi,t)$
is decreasing as a function of $t$. For the sake of comparison, on
the same plot we also plotted in grey lines the nuclear form factors
in conventional exponential parameterization, $F_{A}(t)=\exp\left(\frac{R_{A}^{2}}{6}t\right)$,
and for radius $R_{A}$ we used $R_{A}=1.2\, fm\times A^{1/3}$.
We can see that to a good extent the $t$-dependence of the GPDs is
similar to that of the form factors.

In the right panel of  Figure~\ref{CGC:Results:G_dependence}
we plot the $\xi$-dependence of the gluon GPDs in nuclei. We can
see that in the small-$\xi$ region $H^{g}(x,\xi,t)$ is independent
of the skewedness $\xi$. This results is quite easy to understand:
in the ultrarelativistic limit the nucleus in laboratory frame is
squeezed to an infinitely thin ``pancake\char`\", so the condensate
distribution along the $x^{-}$-axis is strongly peaked around $x^{-}\approx0,\,\lambda_{A}(x^{-})\sim\delta(x^{-}).$
As a consequence, the gluon GPD which is proportional to the Fourier
of $\lambda_{A}(x^{-})$, almost does not depend on $\Delta^{+}\sim\xi$.
The only exception is the region of sufficiently large $\xi\sim0.1$,
where the $\xi$-dependence is mainly a kinematical effect--the increase
of $H^{g}(x,\xi,t)$ is due to decreasing $\Delta_{\perp}$ at fixed
$t$. However, these values of $\xi\sim 0.1$ are too large, and our extrapolation of the model becomes unreliable.

\begin{figure}[h]
\includegraphics[scale=0.3]{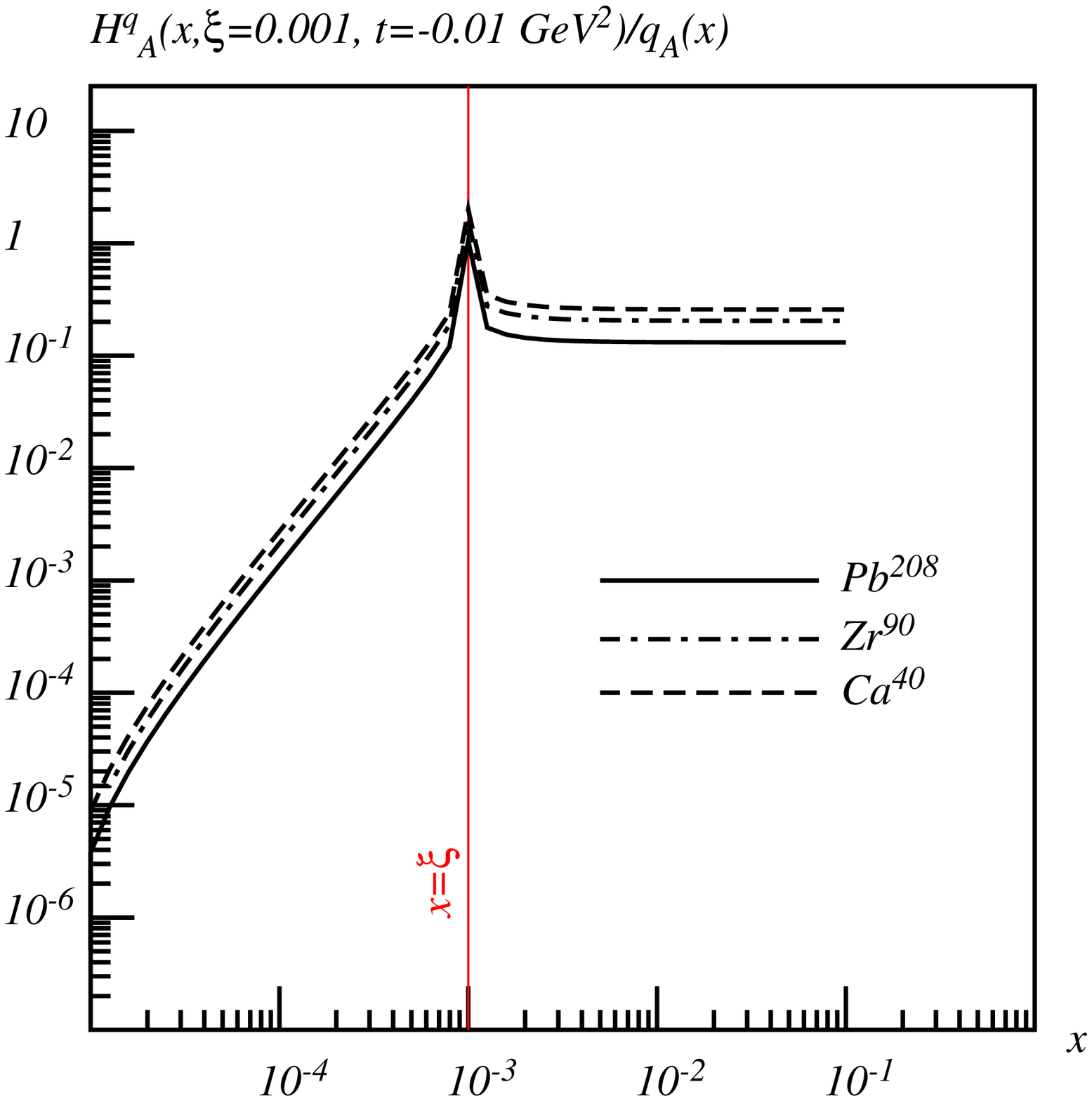} 
\includegraphics[scale=0.3]{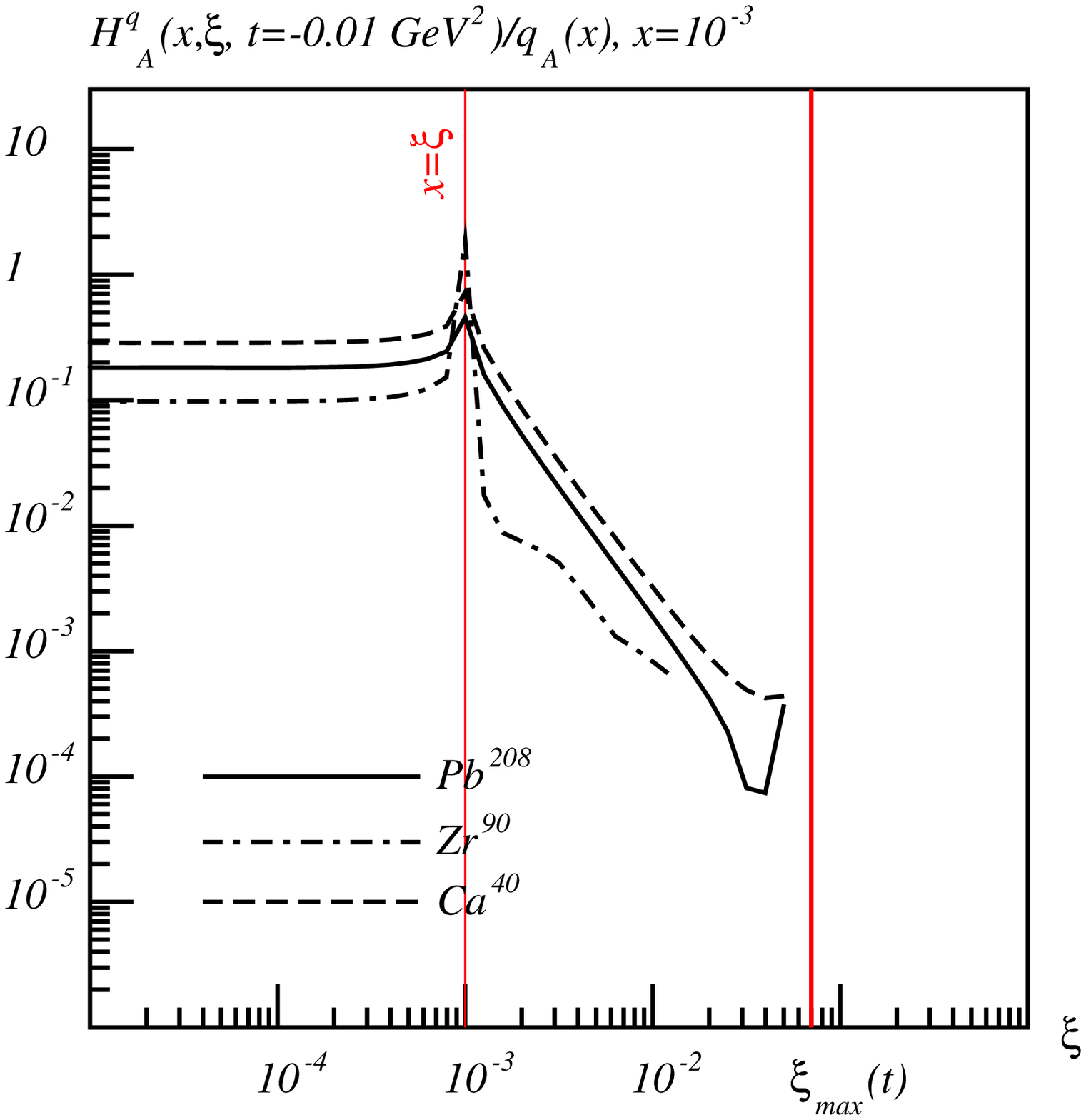}
\includegraphics[scale=0.3]{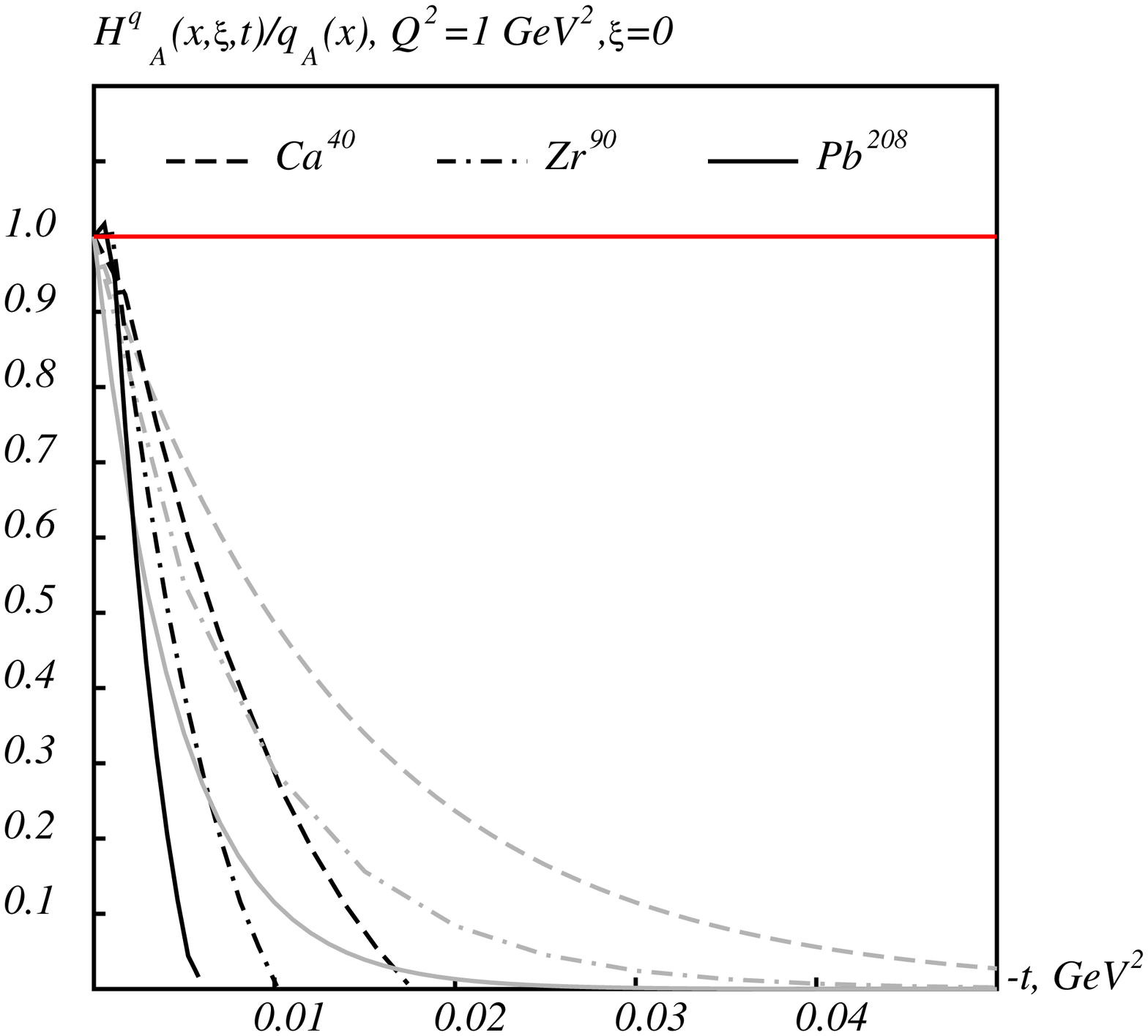}

\caption{\label{CGC:Results:H_dependence} Left plot: $x$-dependence of the
quark GPD $H_{A}(x,\xi,t)$. $\xi=10^{-2},Q^{2}=1\,\GeV^{2}$. Middle
plot: $\xi$-dependence of the same GPD $H_{A}(x,\xi,t)$ for fixed
$x=10^{-4},Q^{2}=1\,\GeV^{2}$,$\xi_{max}=\sqrt{-t/(4M^{2}-t)}$.
Right plot: $t$-dependence of the same GPD $H_{A}(x,\xi,t)$. For
comparison, we also plotted in grey lines the nuclear formfactor in
the the simplest exponential parameterization $F_{A}(t)=\exp\left(\frac{R_{A}^{2}}{6}t\right)$.}

\end{figure}

In Figure~\ref{CGC:Results:H_dependence} we plot the $x$-, $\xi$-
and $t$-dependence of the quark GPD $H_{A}(x,\xi,t)$ in nuclei.
As one can see from~(\ref{CGC:H:Unintegrated:Final}), in the forward
limit the quark distributions have a very simple $x$-dependence,
$H_{A}(x,0,0)\equiv q_{A}(x)\sim1/x.$ For better legibility, we prefer
to discuss out results for the quarks in terms of the ratio $H_{A}(x,\xi,t)/q_{A}(x),$
which measures off-forward effects.

From the left panel in Figure~\ref{CGC:Results:H_dependence} we
can see that for $x\ll\xi$ the GPD $H_{A}(x,\xi,t)$ is decreasing
approximately as $H_{A}(x,\xi,t)\sim x$ and as a result the ratio
$H_{A}(x,\xi,t)/q_{A}(x)$ behaves approximately as $H_{A}(x,\xi,t)/q_{A}(x)\sim x^{2}$.
For $x\gg\xi,$ $H_{A}(x,\xi,t)\approx q_{A}(x)F_{A}(t)$, and the
ratio is a constant. In the point $x=\xi$ we have a singularity $\sim\ln|x-\xi|$,
 which was mentioned
at the end of Section~\ref{sec:Unintegrated-quark-GPDs} and discussed
in details in Appendix~\ref{sec:A:QuarkGPD}.

From the middle panel in Figure~\ref{CGC:Results:H_dependence} we
can see that as a function of $\xi$ the generalized quark distribution
is a constant for $\xi\ll x$, but is a decreasing function for $\xi\gg x$.

From the right panel in Figure~\ref{CGC:Results:H_dependence} we
can see the $t$-dependence of the GPD $H_{A}(x,\xi,t)$. For the
sake of comparison, on the same plot we also plotted in grey lines
the nuclear form factors in the frequently used exponential parameterization,
$F_{A}(t)=\exp\left(\frac{R_{A}^{2}}{6}t\right).$ We can see that
$H_{A}(x,\xi,t)$ is decreasing a bit faster than $F_{A}(t)$.

\begin{figure}[h]
\includegraphics[scale=0.3]{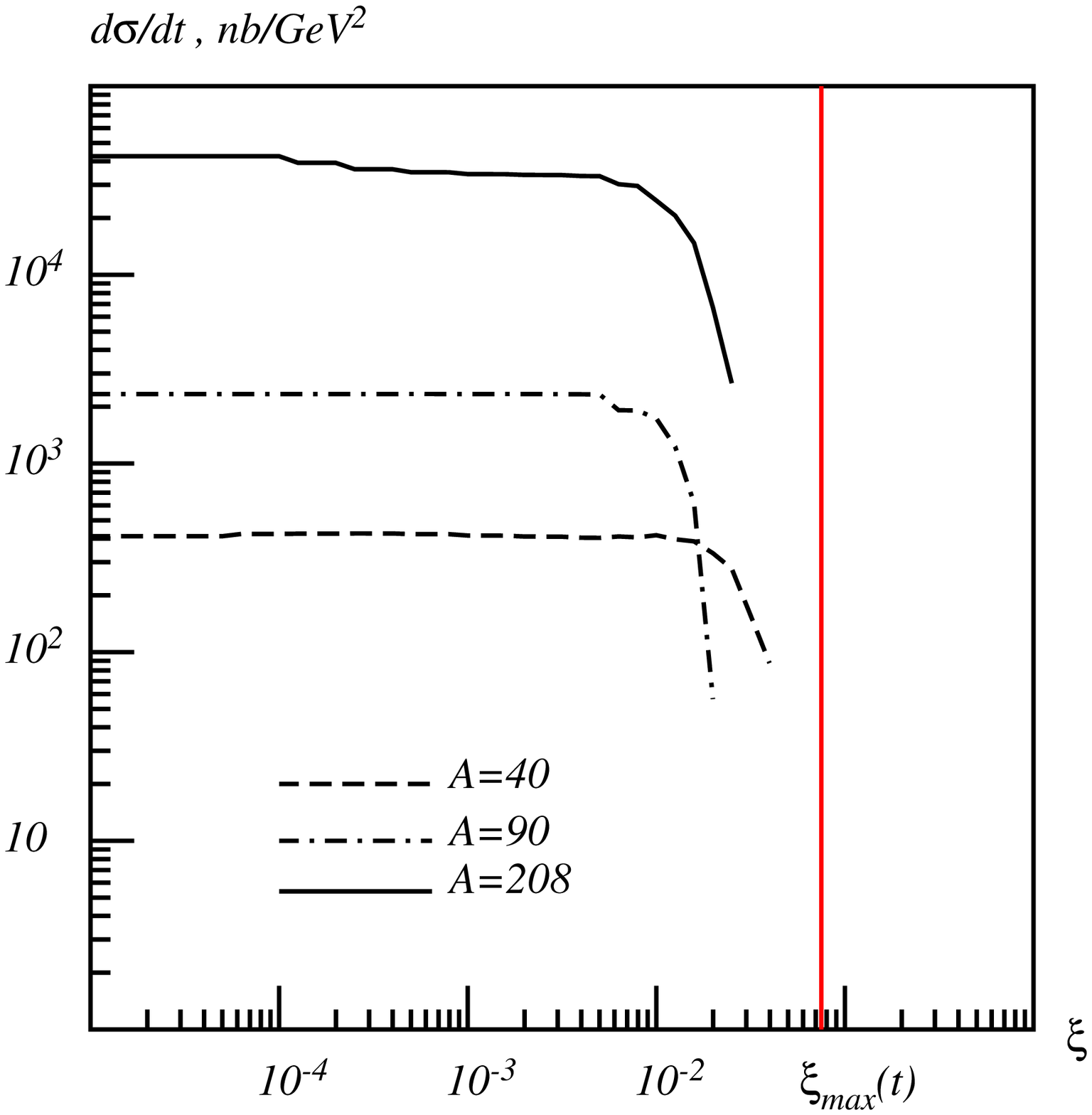} 
\includegraphics[scale=0.3]{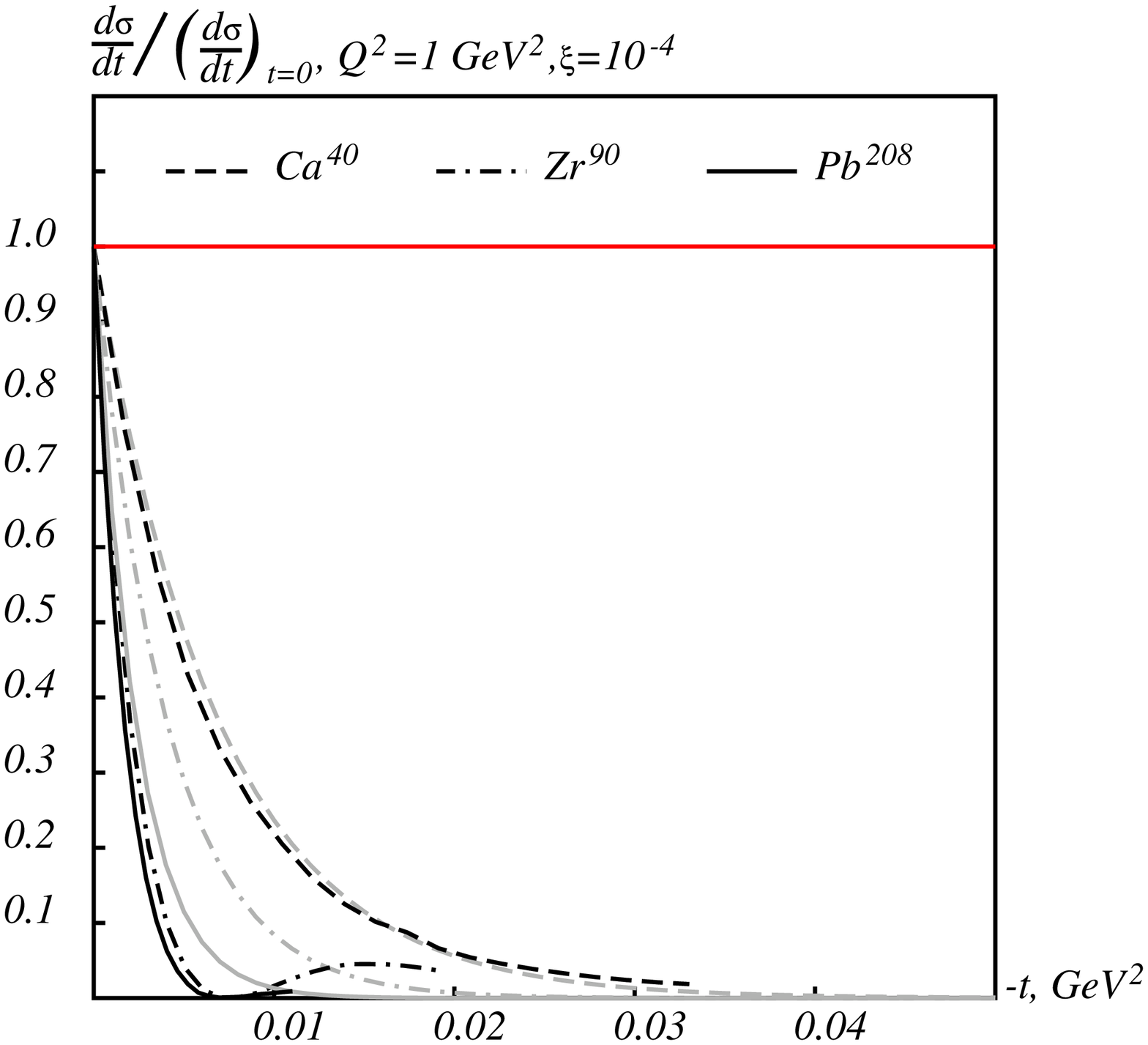}

\caption{\label{CGC:Results:XSection_xi_dependence:nuclei} Left plot: $\xi$-dependence
of the differential DVCS cross-section in the CGC model for different
nuclei. Kinematic is chosen as $Q^{2}=1\,\GeV^{2}$, $t=-0.01\,\GeV^{2}$.
Right plot: $t$-dependence of the DVCS cross-section at fixed $\xi=10^{-4}.$
On the right plot, we also plotted in grey lines what one would have
with the simplest factorized $t$-dependence of the DVCS amplitude
and exponential parameterization for the formfactor: $\frac{d\sigma}{dt}\sim F_{A}^{2}(t)\sim\exp\left(\frac{R_{A}^{2}}{3}t\right)$}

\end{figure}

In Figure~\ref{CGC:Results:XSection_xi_dependence:nuclei} we plot
the $\xi$- and $t$-dependence of the differential DVCS cross-section
$d\sigma/dt$ for fixed $Q^{2}$ and different nuclei.

From the left part of Figure~\ref{CGC:Results:XSection_xi_dependence:nuclei}
we can see that the cross-section is growing when $\xi$ is decreasing,
but at some $\xi$, which we call $\xi_{sat}(Q^{2},A)$, we have a
qualitative transition to the saturation. The value $\xi_{sat}(Q^{2},A)$
depends on the external kinematics. The relatively large value $\xi_{sat}\sim0.01$
is due to the small value of $Q^{2}=1\,\GeV^{2}$, $\xi_{sat}$ is
decreasing when $Q^{2}$ increases.

From the right plot on Figure~\ref{CGC:Results:XSection_xi_dependence:nuclei}
we can see the $t$-dependence of the differential cross-section $d\sigma/dt$.
For the sake of comparison, on the same plot we also plotted in grey
lines the nuclear form factors in conventional exponential parameterization,
$F_{A}(t)=\exp\left(\frac{R_{A}^{2}}{6}t\right).$ We can see that
$d\sigma/dt$ is decreasing a bit faster than $F_{A}^2(t)$.

\subsection{Results with parameterization I}

\label{subsec:Model-II}

In this section we discuss the results of Color Glass Condensate model
in parameterization~(\ref{CGC:Weight:finitewidth}). The crucial
point is that this model explicitly contains the saturation scale
$Q_{s}^{2}$ and as a consequence we can apply it only to the kinematics
where saturation is present. In our evaluations we used for $Q_{s}^{2}$
the parameterization from~\cite{Iancu:2003xm}, where $Q_{s}^{2}(A)$
is found as a solution of the equation \begin{equation}
Q_{s}^{2}(A)=\alpha_{s}(Q^{2})N_{c}\mu_{A}^{2}\ln\left(\frac{Q_{s}^{2}(A)}{\Lambda_{QCD}^{2}}\right).\end{equation}
 This equation has real solutions only for $A\gtrsim A_{min}(Q^{2})\sim150$
for $Q^{2}\sim1\,\GeV^{2},$ and $A_{min}(Q^{2})$ is a growing function
of $Q^{2}.$

\begin{figure}[h]
\includegraphics[scale=0.3]{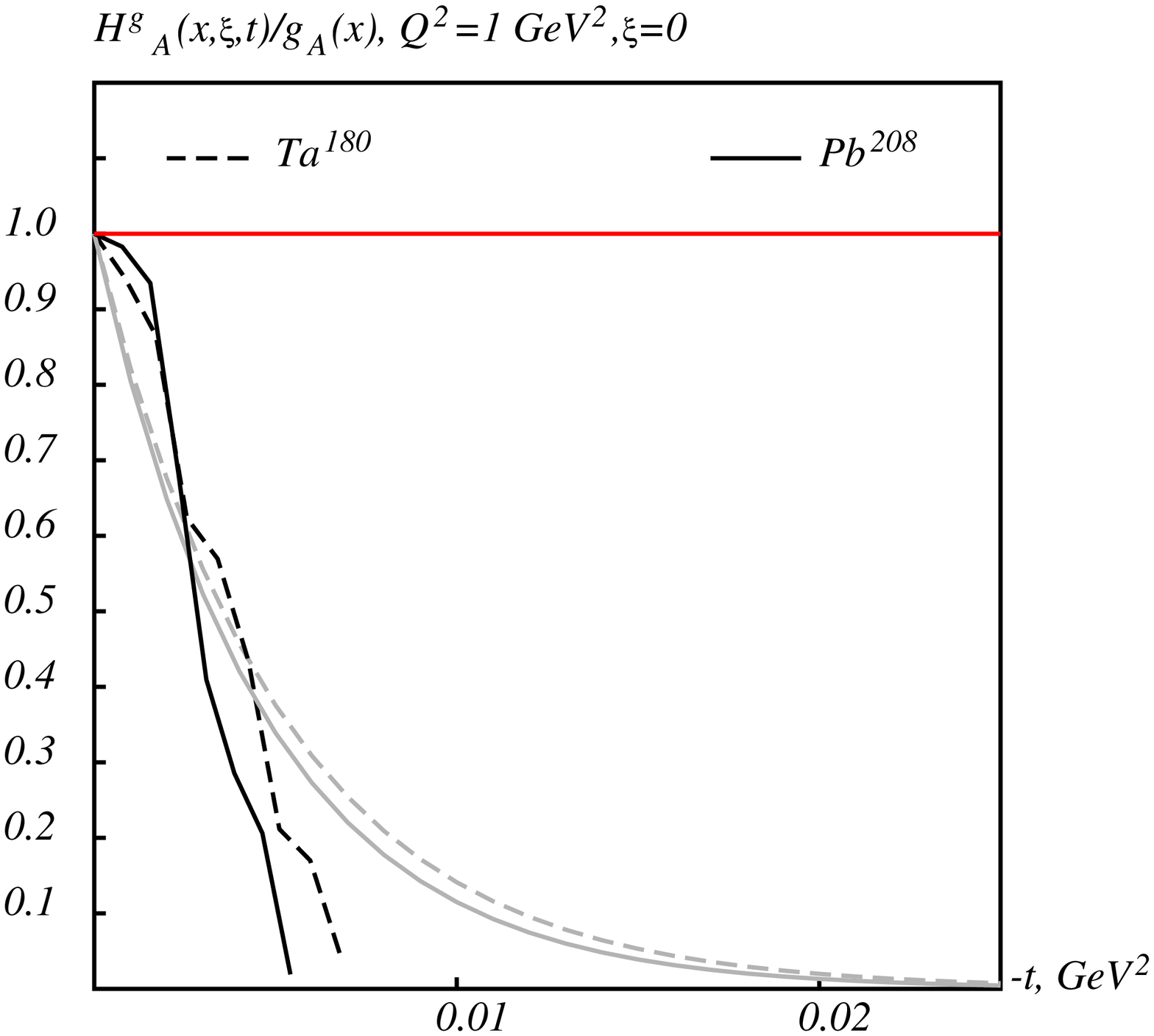} 
\includegraphics[scale=0.3]{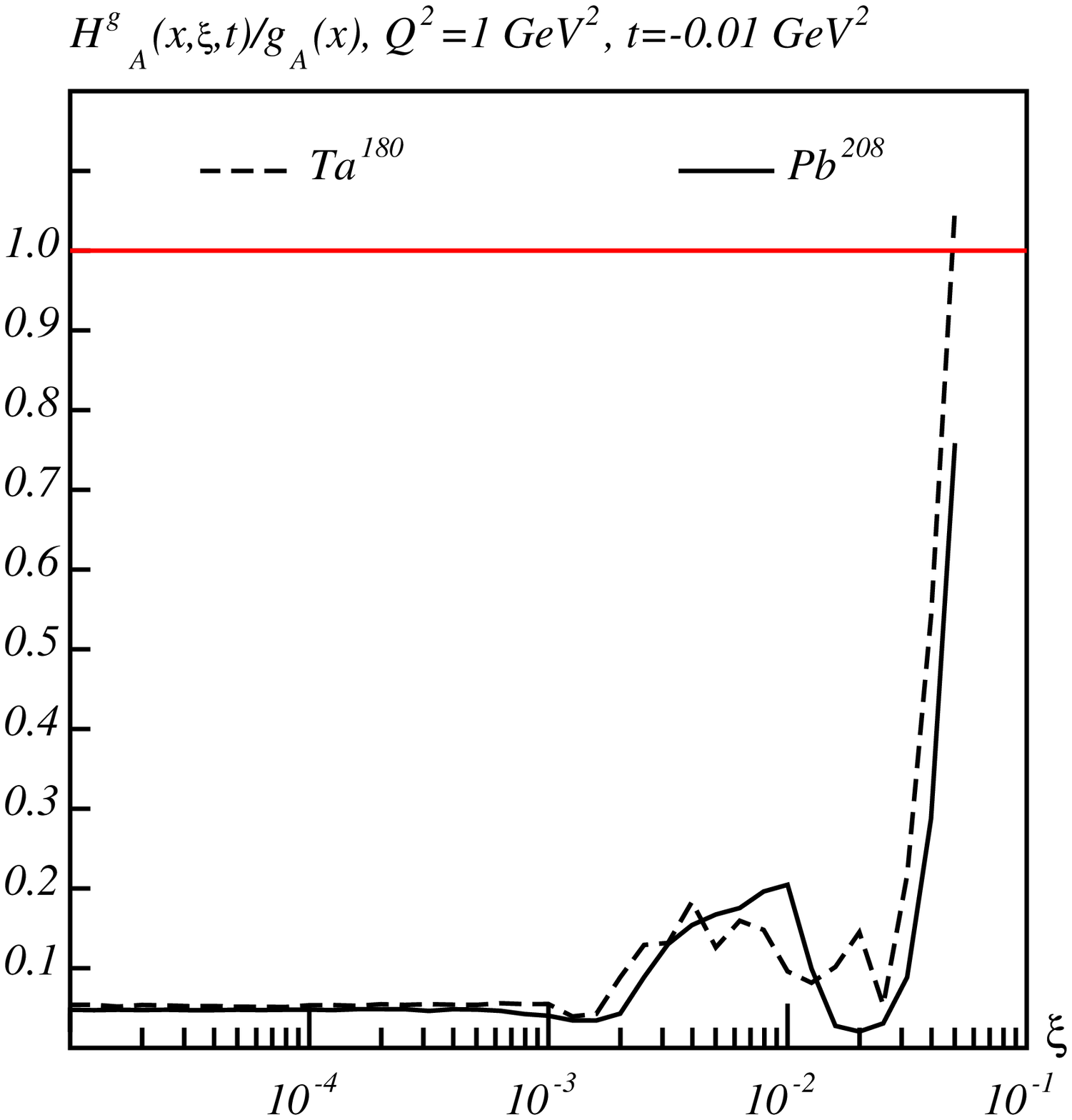}

\caption{\label{CGC:Results:G_dependence-II} Left plot: $t$-dependence of
the gluon distribution for different nuclei. $\xi=0,Q^{2}=1\,\GeV^{2}$.
Right plot: $\xi$-dependence for the same nuclei for fixed $t=-0.01\,\GeV^{2},Q^{2}=1\,\GeV^{2}$.
We do not plot the $x$-dependence of the gluon GPD $H^{g}$, which
is according to~(\ref{eq:GluonGPDFinal}) just a trivial $1/x$ for
all $(\xi,t)$. On the left plot, we also plotted in grey lines the
nuclear formfactor in the the simplest exponential parameterization
$F_{A}(t)=\exp\left(\frac{R_{A}^{2}}{6}t\right)$.}

\end{figure}

\begin{figure}[h]
\includegraphics[scale=0.3]{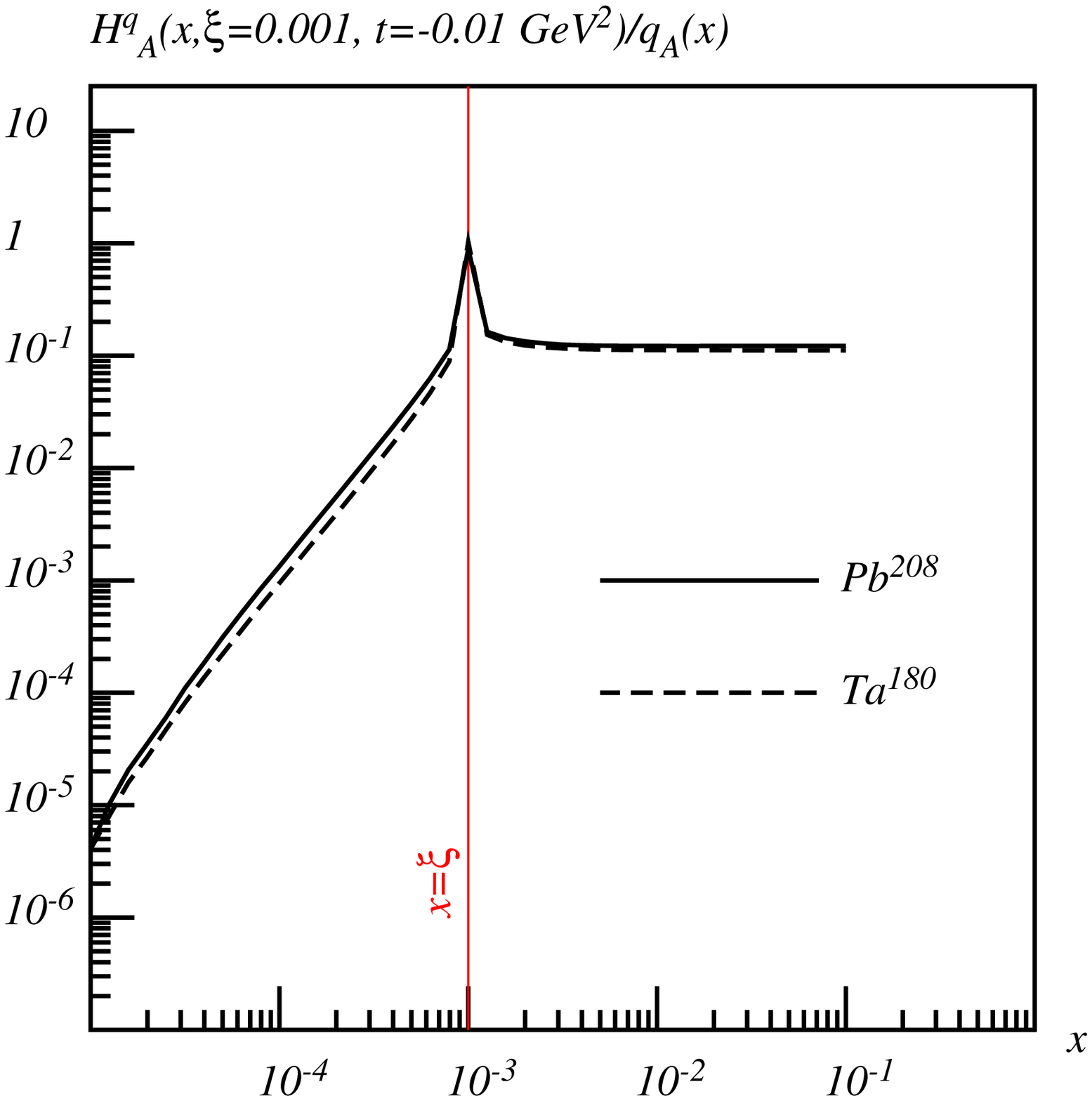} 
\includegraphics[scale=0.3]{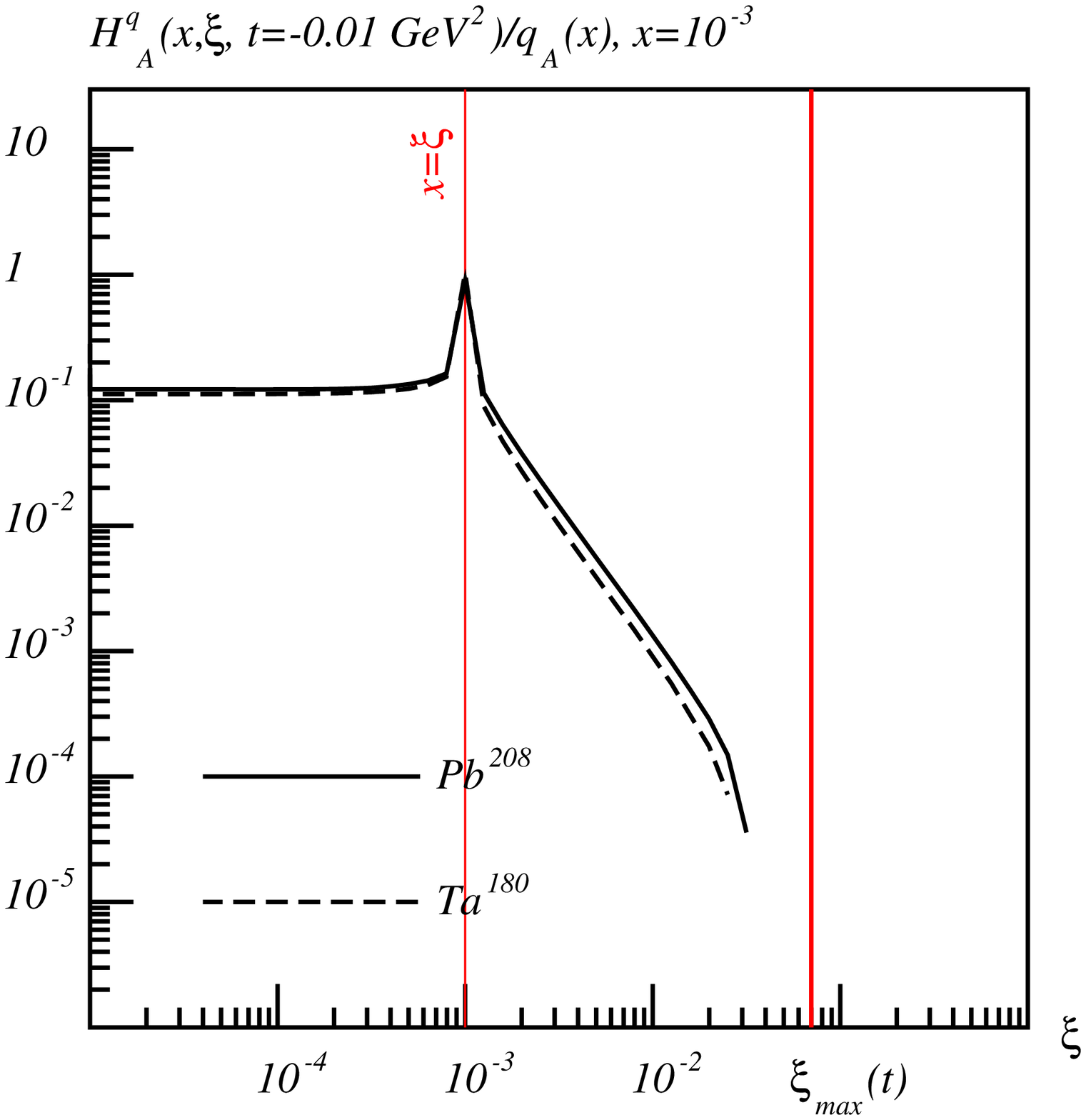}
\includegraphics[scale=0.3]{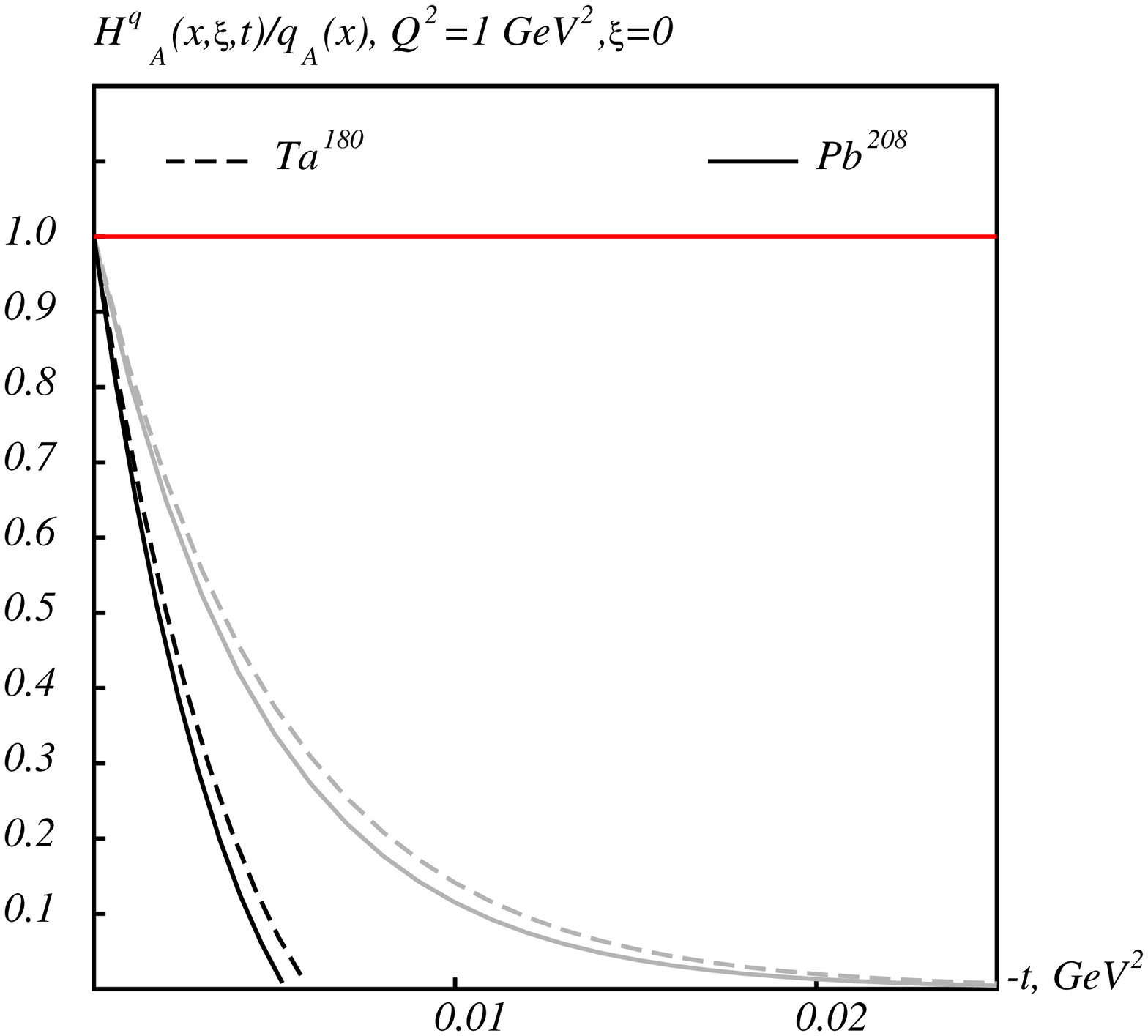}

\caption{\label{CGC:Results:H_dependence-II} Left plot: $x$-dependence of
the quark GPD $H_{A}(x,\xi,t)$. $\xi=10^{-3},Q^{2}=1\,\GeV^{2}$.
Middle plot: $\xi$-dependence of the same GPD $H_{A}(x,\xi,t)$ for
fixed $x=10^{-3},Q^{2}=1\,\GeV^{2}$. Right plot: $t$-dependence
of the same GPD $H_{A}(x,\xi,t)$. On the left plot, we also plotted
in grey lines the nuclear formfactor in the the simplest exponential
parameterization $F_{A}(t)=\exp\left(\frac{R_{A}^{2}}{6}t\right)$.}

\end{figure}

\begin{figure}[h]
\includegraphics[scale=0.3]{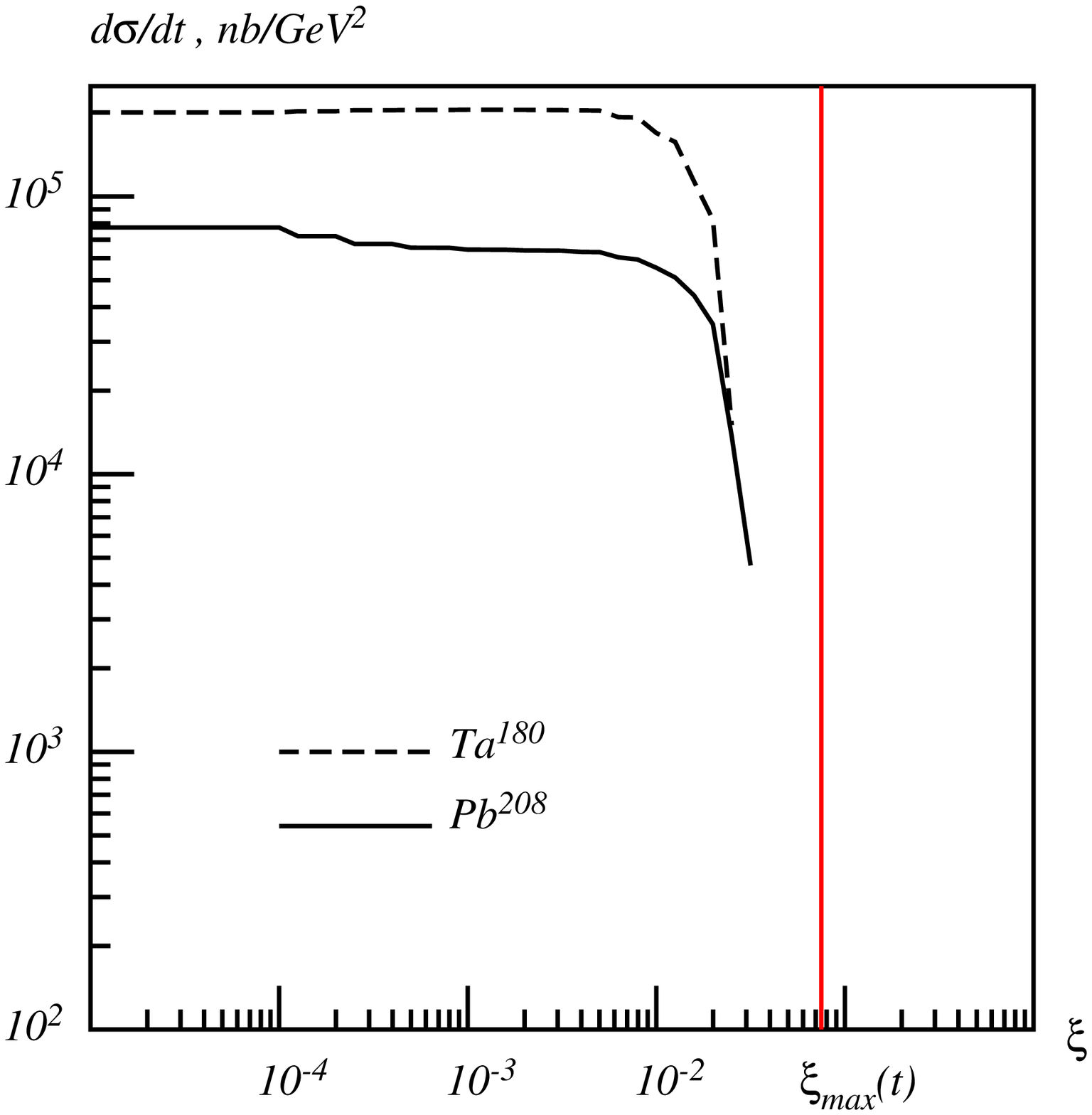} 
\includegraphics[scale=0.3]{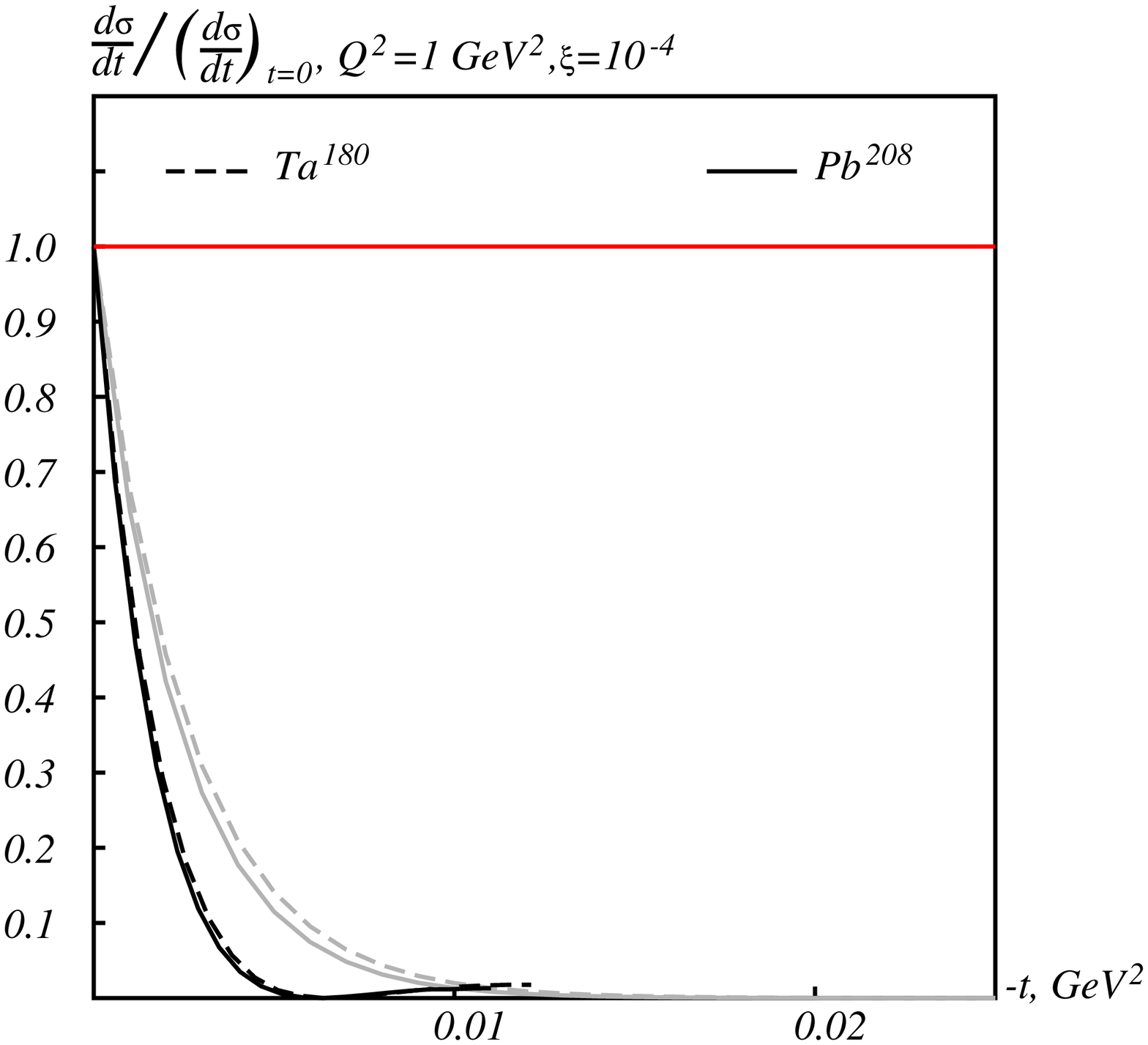}

\caption{\label{CGC:Results:XSection_xi_dependence:nuclei-II} Left plot: $\xi$-dependence
of the differential DVCS cross-section in the CGC model for different
nuclei. Kinematic is chosen as $Q^{2}=1\,\GeV^{2}$, $t=-0.01\,\GeV^{2}$.
Right plot: $t$-dependence of the DVCS cross-section at fixed $\xi=10^{-4}.$
On the right plot, we also plotted in grey lines what one would have
with the simplest factorized $t$-dependence of the DVCS amplitude
and exponential parameterization for the formfactor: $\frac{d\sigma}{dt}\sim F_{A}^{2}(t)\sim\exp\left(\frac{R_{A}^{2}}{3}t\right)$}

\end{figure}

In Figures~\ref{CGC:Results:G_dependence-II} and \ref{CGC:Results:H_dependence-II}
we plot the $\xi$ and $t$-dependence of the gluon and quark distributions
$H_{A}(x,\xi,t)/q_{A}(x),$ and in Figure~\ref{CGC:Results:XSection_xi_dependence:nuclei-II}
we plot the $\xi$- and $t$-dependence of the differential DVCS cross-section
$d\sigma/dt$ . We can see that qualitatively the behaviour is the
same as in the previous section, although absolute values differ.

\subsection{Comparison to DVCS cross section in GVMD model}

\begin{figure}[h]
\includegraphics[scale=0.3]{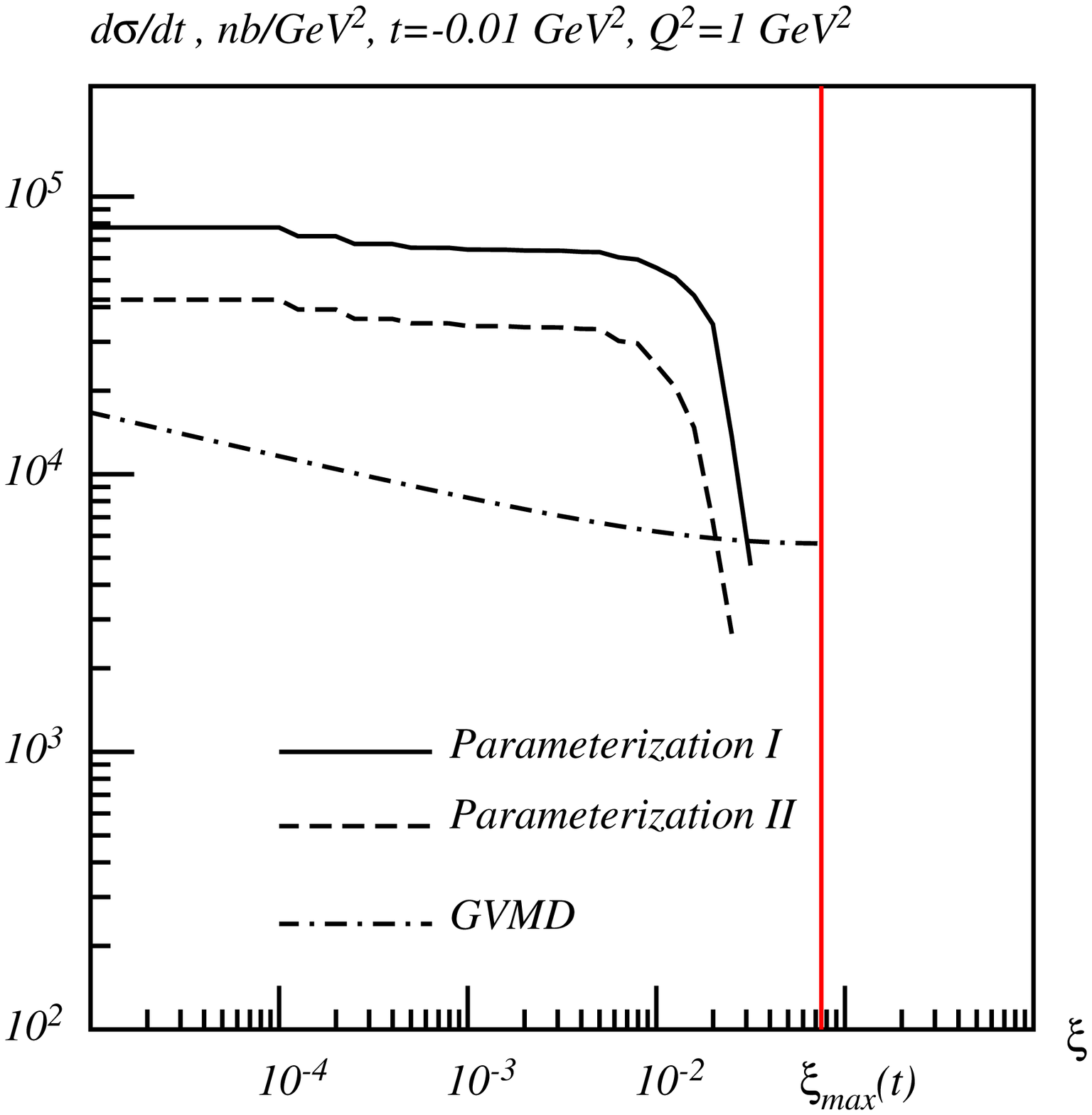}

\caption{\label{CGC:Results:XSection_xi_dependence:models} Comparison of $\xi$-dependence
of the DVCS cross-section in different models. Solid curve corresponds
to parameterization I from Eq.~(\ref{CGC:Weight:finitewidth}), dashed
curve corresponds to parameterization II from Eq.~(\ref{eq:simple-lambdaDefinition}),
dot-dashed corresponds to Generalized Vector Meson Dominance (GVMD)
from~\cite{Goeke:2008rn}. Kinematic is chosen as $Q^{2}=1\,\GeV^{2}$,
$t=-0.01\,\GeV^{2}$, nucleus $A=208$ (lead).}

\end{figure}

In the Figure~\ref{CGC:Results:XSection_xi_dependence:models} we
compare predictions for the DVCS cross-section with our earlier result~\cite{Goeke:2008rn}
obtained in Generalized Vector Dominance Model (GVMD). We can see
the difference in predictions of GVMD and CGC models: In contrast
to the saturation behavior in CGC, the GVMD cross-section is slowly
growing as $\xi^{-\alpha}$ when $\xi$ is decreasing. Nevertheless
in the region $10^{-5}\le\xi\le10^{-3}$ predictions of both models
have comparable values.

\section{Conclusion}

In this paper we considered Generalized Parton Distributions (GPDs)
and Deeply Virtual Compton Scattering (DVCS) amplitudes in the Color
Glass Condensate model. We modified the original formulation of~\cite{McLerran:1993ni,McLerran:1993ka}
to off-forward kinematics of hard exclusive reactions, which provided
the necessary framework for the calculation of GPDs and the DVCS amplitude.

We evaluated the quark and gluon GPDs in this model and studied their
dependence on variables $x,\xi,t$. We found that the gluon GPD $H^{g}$
in this model has a simple $x$-dependence $H^{g}(x)\sim1/x$ for
all $(\xi,t)$. Similar $1/x$-behaviour was observed for the quark
GPDs $H^{q}$ in the $x\gg\xi$ region and in the forward limit ($t=0$).
Both the quark and gluon GPDs are decreasing as a function of momentum
transfer $t$, and the quark GPD is decreasing a bit faster than gluon
GPD.

Without assuming the validity of the collinear factorization, we evaluated
the DVCS cross-sections in the small-$\xi$ region on the large nuclei.
We found that in this region the DVCS cross-sections are almost independent
of $\xi$. This is a manifestation of the general saturation property
inherent to the CGC model. As far as absolute values are concerned,
we found that the predictions of CGC in the relevant range of $\xi$
are comparable with predictions of other models, e.g. GVMD. Currently
there is no experimental data available for DVCS cross-section in
this kinematics.

The present calculation should be important for a wide range of the
future experiments. For example, gluon GPDs in the
small-$x$ region may be used for evaluation of the heavy vector meson production
in ultraperipheral collisions at the LHC~\cite{Baltz:2007kq,Baltz:2007hw}.

\section*{Acknowledgments}

We would like to thank P. Pobylitsa and M. Strikman for useful discussions.
The work has been partially supported by the Collaborative Research
Center Bonn-Bochum-Giessen of the DFG, by the I3HP European Project
(6-th Framework), by the Verbundforschung ``Hadrons and Nuclei''
of the BMBF, by the Graduate College Dortmund-Bochum of the DFG, by
the COSY-Project Juelich, and by the AvH-Kovalevskaja Funds (M.Polyakov).

\appendix
\textbf{Notice:} Authored by Jefferson Science Associates, LLC under
U.S. DOE Contract No. DE-AC05-06OR23177. The U.S. Government retains
a non-exclusive, paid-up, irrevocable, world-wide license to publish
or reproduce this manuscript for U.S. Government purposes.

\section{Details of evaluation of (\ref{CGC:H:Unintegrated:quasiclassical})}

\label{sec:A:QuarkGPD} In this section we evaluate the unintegrated
GPD~(\ref{CGC:H:Unintegrated}), which in quasiclassical approximation
was reduced to 
\begin{eqnarray}
&  & H(x,\xi,\vec{\Delta}_{\perp},\vec{k}_{\perp})=\nonumber\\
&  & =i\bar{P}^{+}\int\frac{dz^{-}}{2\pi}\int d^{2}r_{\perp}e^{-i\vec{k}_{\perp}\vec{r}_{\perp}}\int d^{3}Xe^{-i\vec{\Delta}\vec{X}}\left\langle Tr\left[\gamma^{+}S\left(-\frac{z^{-}}{2}-\frac{\vec{r}_{\perp}}{2}-\vec{X},\frac{z^{-}}{2}+\frac{\vec{r}_{\perp}}{2}-\vec{X}\right)\right]\right\rangle \nonumber \\
&  &= i\bar{P}^{+}\int\frac{d^{3}\xi_{1}}{(2\pi)^{3}}\frac{d^{3}\xi_{2}}{(2\pi)^{3}}e^{ip_{2}\cdot\xi_{2}-ip_{1}\cdot\xi_{1}}\left\langle Tr\left[\gamma^{+}S(\xi_{1};\xi_{2})\right]\right\rangle \,,
\end{eqnarray}

where we changed the integration variables according to
\begin{eqnarray}
\vec{\xi}_{1}=-\frac{z^{-}}{2}-\frac{\vec{r}_{\perp}}{2}-\vec{X}\,,\\
\vec{\xi}_{2}=\frac{z^{-}}{2}+\frac{\vec{r}_{\perp}}{2}-\vec{X}\,,
\end{eqnarray}
and introduced shorthand notations
\begin{eqnarray}
\vec{p}_{1}=x\bar{P}^{+}+\vec{k}_{\perp}-\frac{\vec{\Delta}}{2}\,,\\
\vec{p}_{2}=x\bar{P}^{+}+\vec{k}_{\perp}+\frac{\vec{\Delta}}{2}\,.
\end{eqnarray}
 
Now we have to consider separately the first case $\xi_{1}^{-}>0,\xi_{2}^{-}<0$
and the second case $\xi_{1}^{-}<0,\xi_{2}^{-}>0$. All the other
regions are just the vacuum contributions $\sim\delta^{2}\left(\Delta_{\perp}\right)$
and must be omitted. For the sake of brevity we will refer to the
contribution of the first region as $H^{+-}$, and to the second one
as $H^{-+}$.

For $\theta$-functions of arguments $\pm\xi_{1,2}$ we will use an
integral representation \begin{eqnarray}
\theta(\pm\xi)=\frac{1}{2\pi i}\int_{-\infty}^{\infty}d\alpha\frac{e^{\pm i\alpha\xi}}{\alpha-i0}=-\frac{1}{2\pi i}\int_{-\infty}^{\infty}d\alpha\frac{e^{\mp i\alpha\xi}}{\alpha+i0}.\label{CGCF:theta}\end{eqnarray}

\subsection{Evaluation of $H^{+-}$}

In the first case we have explicitly

\begin{eqnarray}
 &  & H^{+-}=-i\bar{P}^{+}\int\frac{d^{3}\xi_{1}}{(2\pi)^{3}}\frac{d^{3}\xi_{2}}{(2\pi)^{3}}e^{ip_{2}\cdot\xi_{2}-ip_{1}\cdot\xi_{1}}\int\frac{d\alpha_{1}d\alpha_{2}}{(2\pi)^{2}}\frac{e^{i(\alpha_{1}\xi_{1}^{-}-\alpha_{2}\xi_{2}^{-})}}{(\alpha_{1}-i0)(\alpha_{2}-i0)}\nonumber \\
 &  & \times\int\frac{d^{4}p}{(2\pi)^{4}}\frac{1}{p^{2}-M^{2}+i0}\int\frac{d^{2}q}{(2\pi)^{2}}\exp\left\{ -i\left(\frac{q_{\perp}^{2}+M^{2}}{2p^{-}}\xi_{1}^{-}-q_{\perp}\cdot\xi_{1\perp}-p^{+}\xi_{2}^{-}+\vec{p}_{\perp}\xi_{2}^{\perp}\right)\right\}\nonumber \\
 &  & \times\int d^{2}z\, e^{i(p_{\perp}-q_{\perp})z}Tr\left[\gamma_{+}\left(1+\frac{\gamma_{0}}{p^{-}\sqrt{2}}(\hat{q}_{\perp}+M)\right)\Lambda^{(-)}(\hat{p}+M)\right]\left\langle U^{\dagger}(z)U(\xi_{1\perp})\right\rangle \,.\end{eqnarray}

Now  evaluate each of the integrals:

\begin{eqnarray}
&  & \int\frac{d\xi_{1}^{-}d\xi_{2}^{-}}{(2\pi)^{2}}e^{ip_{1}^{+}\xi_{1}^{-}-ip_{2}^{+}\xi_{2}^{-}}e^{i(\alpha_{1}\xi_{1}^{-}-\alpha_{2}\xi_{2}^{-})}\exp\left(-i\frac{q_{\perp}^{2}+M^{2}}{2p^{-}}\xi_{1}^{-}+ip^{+}\xi_{2}^{-}\right)\nonumber\\
&  &= \delta\left(\alpha_{1}+p_{1}^{+}-\frac{q_{\perp}^{2}+M^{2}}{2p^{-}}\right)\delta\left(\alpha_{2}+p_{2}^{+}-p^{+}\right)\,, \\
&  & \int\frac{d^{2}p_{\perp}}{(2\pi)^{2}}\int d^{2}z\, e^{i(p_{\perp}-q_{\perp})z}\int\frac{d^{2}\xi_{1}^{\perp}d^{2}\xi_{2}^{\perp}}{(2\pi)^{4}}e^{-ip_{1}^{\perp}\xi_{1}^{\perp}+ip_{2}^{\perp}\xi_{2}^{\perp}}e^{iq_{\perp}\xi_{1}^{\perp}-ip_{\perp}\xi_{2}^{\perp}}\left\langle U(\xi_{1\perp})U^{\dagger}(z)\right\rangle \nonumber\\
&  & =\int d^{2}z\, e^{i(p_{2\perp}-q_{\perp})z}\int\frac{d^{2}\xi_{1}^{\perp}}{(2\pi)^{2}}e^{-i(p_{1}^{\perp}-q_{\perp})\xi_{1}^{\perp}}\left\langle U^{\dagger}(z)U(\xi_{1\perp})\right\rangle \left|_{p_{\perp}=p_{2\perp}}\right.\,. \end{eqnarray}
 Now change the dummy integration variables $\vec{\xi}_{1}^{\perp}$
and $\vec{z}$ to $\vec{X}^{\perp}$ and $\vec{\rho}^{\perp}$: \begin{eqnarray}
 &  & \xi_{1}^{\perp}:=X^{\perp}-\frac{\rho^{\perp}}{2}\,,\\
 &  & z:=X^{\perp}+\frac{\rho^{\perp}}{2}\nonumber \,,\\
 &  & \Rightarrow\int d^{2}z\, e^{i(p_{2\perp}-q_{\perp})z}\int\frac{d^{2}\xi_{1}^{\perp}d^{2}\xi_{2}^{\perp}}{(2\pi)^{4}}e^{-ip_{1}^{\perp}\xi_{1}^{\perp}+ip_{2}^{\perp}\xi_{2}^{\perp}}e^{iq_{\perp}\xi_{1}^{\perp}-ip_{2\perp}\xi_{2}^{\perp}}\left\langle U(\xi_{1\perp})U^{\dagger}(z)\right\rangle \nonumber\\
 &  &= \tilde{\gamma}\left(k^{\perp}-q^{\perp}+\frac{\Delta^{\perp}}{2},k^{\perp}-q^{\perp}-\frac{\Delta^{\perp}}{2}\right)\,, \end{eqnarray}
 where $\tilde{\gamma}$ was defined in~(\ref{tildeGamma:definition}).
It is convenient to make a shift of the dummy integration variable
according to \begin{equation}
\int\frac{d^{2}q^{\perp}}{(2\pi)^{2}}\to\int\frac{d^{2}\kappa^{\perp}}{(2\pi)^{2}}\qquad\mbox{where}\qquad\vec{\kappa}^{\perp}=\vec{k}^{\perp}-\vec{q}^{\perp}\,.\end{equation}

\begin{eqnarray}
 &  & \Rightarrow H^{+-}(x,\xi,t,k_{\perp})=-iN_{c}\int\frac{d^{2}\kappa^{\perp}}{(2\pi)^{2}}\tilde{\gamma}\left(\kappa^{\perp}+\frac{\Delta^{\perp}}{2},\kappa^{\perp}-\frac{\Delta^{\perp}}{2}\right)\int\frac{dp^{+}dp^{-}}{(2\pi)^{2}}\frac{1}{2p^{+}p^{-}-p_{2}^{\perp2}-M^{2}+i0}\nonumber\\
 &  &\times \frac{1}{p^{+}-p_{2}^{+}-i0}\,\frac{1}{\frac{(k_{\perp}-\kappa_{\perp})^{2}+M^{2}}{2p^{-}}-p_{1}^{+}-i0}\,\frac{M^{2}-\vec{p}_{2\perp}(\vec{k}_{\perp}-\vec{\kappa}_{\perp})}{p^{-}}\label{eq:H+-:1} \end{eqnarray}

Now we take the integrals over $p^+\,,p^-$
in~(\ref{eq:H+-:1}). The first integral is taken over $p^{+},$
the result is

\begin{equation}
\int\frac{dp^{+}}{(2\pi)}\frac{1}{2p^{+}p^{-}-p_{2}^{\perp2}-M^{2}+i0}\frac{1}{p^{+}-p_{2}^{+}-i0}=\frac{i\theta(p^{-})}{2p_{2}^{+}p^{-}-(p_{2}^{\perp})^{2}-M^{2}+i0}\,.
\end{equation} 
Integration over $p^-$ yields
\begin{eqnarray}
 &  & H^{+-}(x,\xi,t,k_{\perp})=2\, N_{c}\int\frac{d^{2}\kappa^{\perp}}{(2\pi)^{2}}\tilde{\gamma}\left(\kappa^{\perp}+\frac{\Delta^{\perp}}{2},\kappa^{\perp}-\frac{\Delta^{\perp}}{2}\right)\nonumber \\
 &  & \times\frac{M^{2}-\vec{p}_{2\perp}\cdot(\vec{k}_{\perp}-\vec{\kappa}_{\perp})}{(x-\xi)((\vec{k}_{\perp}-\vec{\kappa}_{\perp})^{2}+M^{2})-(x+\xi)(p_{2}^{\perp2}+M^{2})}\ln\left|\frac{x-\xi}{x+\xi}\frac{((\vec{k}_{\perp}-\vec{\kappa}_{\perp})^{2}+M^{2})}{(p_{2}^{\perp2}+M^{2})}\right|\nonumber \\
 &  & =2\, N_{c}\int\frac{d^{2}\kappa^{\perp}}{(2\pi)^{2}}\tilde{\gamma}\left(\kappa^{\perp}+\frac{\Delta^{\perp}}{2},\kappa^{\perp}-\frac{\Delta^{\perp}}{2}\right)\nonumber\\
 &  & \times\frac{M^{2}-\left(\vec{k}+\frac{\vec{\Delta}_{\perp}}{2}\right)\cdot\left(\vec{k}-\vec{\kappa}_{\perp}\right)}{(x-\xi)\left(\left(\vec{k}-\vec{\kappa}_{\perp}\right)^{2}+M^{2}\right)-(x+\xi)\left(\left(\vec{k}+\frac{\vec{\Delta}_{\perp}}{2}\right)^{2}+M^{2}\right)}\ln\left|\frac{x-\xi}{x+\xi}\frac{\left(\vec{k}-\vec{\kappa}_{\perp}\right)^{2}+M^{2}}{\left(\vec{k}+\frac{\vec{\Delta}_{\perp}}{2}\right)^{2}+M^{2}}\right|\,.\label{eq:A:H+-} \end{eqnarray}

\subsection{Evaluation of $H^{-+}$}

In complete analogy we evaluate the term $H^{-+}$:

\begin{eqnarray}
 &  & H^{-+}=-i\int\frac{d^{3}\xi_{1}}{(2\pi)^{3}}\frac{d^{3}\xi_{2}}{(2\pi)^{3}}e^{ip_{2}\cdot\xi_{2}-ip_{1}\cdot\xi_{1}}\int\frac{d\alpha_{1}d\alpha_{2}}{(2\pi)^{2}}\frac{e^{i(\alpha_{1}\xi_{1}^{-}-\alpha_{2}\xi_{2}^{-})}}{(\alpha_{1}+i0)(\alpha_{2}+i0)}\nonumber\\
 &  &\times \int\frac{d^{4}p}{(2\pi)^{4}}\frac{1}{p^{2}-M^{2}+i0}\int\frac{d^{2}q}{(2\pi)^{2}}\exp\left\{ i\left(\frac{q_{\perp}^{2}+M^{2}}{2p^{-}}\xi_{2}^{-}-q_{\perp}\cdot\xi_{2\perp}-p^{+}\xi_{1}^{-}+\vec{p}_{\perp}\xi_{1}^{\perp}\right)\right\} \nonumber \\
 &  &\times \int d^{2}z\, e^{-i(p_{\perp}-q_{\perp})z}Tr\left[\gamma_{+}(\hat{p}+M)\Lambda^{(+)}\left(1+\frac{\gamma_{0}}{p^{-}\sqrt{2}}(M-\hat{k}_{\perp})\right)\right]\left\langle U^{\dagger}(\xi_{2}^{\perp})U(z)\right\rangle  \end{eqnarray}

Now take the integrals term-by-term in complete analogy with the previous
case

\begin{eqnarray}
 &  & \int\frac{d\xi_{1}^{-}d\xi_{2}^{-}}{(2\pi)^{2}}e^{ip_{1}^{+}\xi_{1}^{-}-ip_{2}^{+}\xi_{2}^{-}}e^{i(\alpha_{1}\xi_{1}^{-}-\alpha_{2}\xi_{2}^{-})}\exp\left(i\frac{q_{\perp}^{2}+M^{2}}{2p^{-}}\xi_{2}^{-}-ip^{+}\xi_{1}^{-}\right)\nonumber\\
 &  & =\delta\left(\alpha_{1}+p_{1}^{+}-p^{+}\right)\delta\left(\alpha_{2}+p_{2}^{+}-\frac{q_{\perp}^{2}+M^{2}}{2p^{-}}\right)\,, \\
 &  & \int\frac{d^{2}p_{\perp}}{(2\pi)^{2}}\int d^{2}z\, e^{-i(p_{\perp}-q_{\perp})z}\int\frac{d^{2}\xi_{1}^{\perp}d^{2}\xi_{2}^{\perp}}{(2\pi)^{4}}e^{-ip_{1}^{\perp}\xi_{1}^{\perp}+ip_{2}^{\perp}\xi_{2}^{\perp}}e^{-iq_{\perp}\xi_{2}^{\perp}+ip_{\perp}\xi_{1}^{\perp}}\left\langle U^{\dagger}(\xi_{2}^{\perp})U(z)\right\rangle \nonumber \\
 &  & =\int d^{2}z\, e^{-i(p_{1\perp}-q_{\perp})z}\int\frac{d^{2}\xi_{2}^{\perp}}{(2\pi)^{2}}e^{i(p_{2}^{\perp}-q_{\perp})\xi_{2}^{\perp}}\left\langle U^{\dagger}(\xi_{2\perp})U(z)\right\rangle \left|_{p_{\perp}=p_{1\perp}}\right.\,, \end{eqnarray}
 Now change the dummy integration variables $\vec{\xi}_{2}^{\perp},\vec{z}$
to $\vec{X},\vec{\rho}^{\perp}$ according to 
\begin{eqnarray}
&  & \xi_{2}^{\perp}:=X^{\perp}+\frac{\rho^{\perp}}{2}\,,\\
&  & z:=X^{\perp}-\frac{\rho^{\perp}}{2}\,.
\end{eqnarray}

\begin{eqnarray}
 &  & \Rightarrow\int\frac{d^{2}p_{\perp}}{(2\pi)^{2}}\int d^{2}z\, e^{-i(p_{\perp}-q_{\perp})z}\int\frac{d^{2}\xi_{1}^{\perp}d^{2}\xi_{2}^{\perp}}{(2\pi)^{4}}e^{-ip_{1}^{\perp}\xi_{1}^{\perp}+ip_{2}^{\perp}\xi_{2}^{\perp}}e^{-iq_{\perp}\xi_{2}^{\perp}+ip_{\perp}\xi_{1}^{\perp}}\left\langle U^{\dagger}(\xi_{2}^{\perp})U(z)\right\rangle \nonumber\\
 &  &= \tilde{\gamma}\left(k^{\perp}-q^{\perp}+\frac{\Delta^{\perp}}{2},k^{\perp}-q^{\perp}-\frac{\Delta^{\perp}}{2}\right) \end{eqnarray}
 where function $\tilde{\gamma}$ was defined in~(\ref{tildeGamma:definition}).
Now shift the dummy integration variable according to \begin{eqnarray}
 &  & \int\frac{d^{2}q^{\perp}}{(2\pi)^{2}}\to\int\frac{d^{2}\kappa^{\perp}}{(2\pi)^{2}}\qquad\mbox{where}\qquad\vec{\kappa}^{\perp}=\vec{k}_{\perp}-\vec{q}^{\perp}\nonumber \\
 &  & \Rightarrow H^{-+}(x,\xi,t,k_{\perp})=+iN_{c}\int\frac{d^{2}\kappa^{\perp}}{(2\pi)^{2}}\tilde{\gamma}\left(\kappa^{\perp}+\frac{\Delta^{\perp}}{2},\kappa^{\perp}-\frac{\Delta^{\perp}}{2}\right)\int\frac{dp^{+}dp^{-}}{(2\pi)^{2}}\frac{1}{2p^{+}p^{-}-p_{1}^{\perp2}-M^{2}+i0}\nonumber\\
 &  &\times \frac{1}{p^{+}-p_{1}^{+}+i0}\,\frac{1}{\frac{(k_{\perp}-\kappa_{\perp})^{2}+M^{2}}{2p^{-}}-p_{2}^{+}+i0}\,\frac{M^{2}-\vec{p}_{1\perp}(\vec{k}_{\perp}-\vec{\kappa}_{\perp})}{p^{-}}\,.\label{eq:H-+:1} \end{eqnarray}

First take the integral over the $p^{+}:$

\begin{equation}
\int\frac{dp^{+}}{(2\pi)}\,\frac{1}{2p^{+}p^{-}-p_{1}^{\perp2}-M^{2}+i0}\,\frac{1}{p^{+}-p_{1}^{+}+i0}=-\frac{i\theta(-p^{-})}{2p_{1}^{+}p^{-}-(p_{1}^{\perp})^{2}-M^{2}+i0}\,,
\end{equation}

next take the integral over $p^{-}:$

\begin{eqnarray}
 & \Rightarrow & H^{-+}=2\, N_{c}\int\frac{d^{2}\kappa^{\perp}}{(2\pi)^{2}}\tilde{\gamma}\left(\kappa^{\perp}+\frac{\Delta^{\perp}}{2},\kappa^{\perp}-\frac{\Delta^{\perp}}{2}\right)\nonumber \\
 &  & \times\frac{M^{2}-\vec{p}_{1\perp}\cdot(\vec{k}_{\perp}-\vec{\kappa}_{\perp})}{(x+\xi)((\vec{k}_{\perp}-\vec{\kappa}_{\perp})^{2}+M^{2})-(x-\xi)((\vec{p}_{1}^{\perp})^{2}+M^{2})}\ln\left|\frac{x+\xi}{x-\xi}\frac{(\vec{k}_{\perp}-\vec{\kappa}_{\perp})^{2}+M^{2}}{(\vec{p}_{1}^{\perp})^{2}+M^{2}}\right|\nonumber \\
 & = & 2\, N_{c}\int\frac{d^{2}\kappa^{\perp}}{(2\pi)^{2}}\tilde{\gamma}\left(\kappa^{\perp}+\frac{\Delta^{\perp}}{2},\kappa^{\perp}-\frac{\Delta^{\perp}}{2}\right)\nonumber\\
 &  & \times\frac{M^{2}-\left(\vec{k}-\frac{\vec{\Delta}}{2}\right)\cdot(\vec{k}_{\perp}-\vec{\kappa}_{\perp})}{(x+\xi)\left(\left(\vec{k}_{\perp}-\vec{\kappa}_{\perp}\right)^{2}+M^{2}\right)-(x-\xi)\left(\left(\vec{k}-\frac{\vec{\Delta}}{2}\right)^{2}+M^{2}\right)}\ln\left|\frac{x+\xi}{x-\xi}\frac{(\vec{k}_{\perp}-\vec{\kappa}_{\perp})^{2}+M^{2}}{\left(\vec{k}-\frac{\vec{\Delta}}{2}\right)^{2}+M^{2}}\right|\,. \end{eqnarray}

In summary, we have

\begin{eqnarray}
H^{+-} & = & 2\, N_{c}\int\frac{d^{2}\kappa^{\perp}}{(2\pi)^{2}}\tilde{\gamma}\left(\kappa^{\perp}+\frac{\Delta^{\perp}}{2},\kappa^{\perp}-\frac{\Delta^{\perp}}{2}\right)\nonumber\\
&  & \times\frac{M^{2}-\left(\vec{k}+\frac{\vec{\Delta}_{\perp}}{2}\right)\cdot\left(\vec{k}-\vec{\kappa}_{\perp}\right)}{(x-\xi)\left(\left(\vec{k}-\vec{\kappa}_{\perp}\right)^{2}+M^{2}\right)-(x+\xi)\left(\left(\vec{k}+\frac{\vec{\Delta}_{\perp}}{2}\right)^{2}+M^{2}\right)}\ln\left|\frac{x-\xi}{x+\xi}\,\frac{\left(\vec{k}-\vec{\kappa}_{\perp}\right)^{2}+M^{2}}{\left(\vec{k}+\frac{\vec{\Delta}_{\perp}}{2}\right)^{2}+M^{2}}\right|\,, \\
H^{-+} & = & 2\, N_{c}\int\frac{d^{2}\kappa^{\perp}}{(2\pi)^{2}}\tilde{\gamma}\left(\kappa^{\perp}+\frac{\Delta^{\perp}}{2},\kappa^{\perp}-\frac{\Delta^{\perp}}{2}\right)\nonumber\\
&  & \times\frac{M^{2}-\left(\vec{k}-\frac{\vec{\Delta}}{2}\right)\cdot(\vec{k}_{\perp}-\vec{\kappa}_{\perp})}{(x+\xi)\left(\left(\vec{k}_{\perp}-\vec{\kappa}_{\perp}\right)^{2}+M^{2}\right)-(x-\xi)\left(\left(\vec{k}-\frac{\vec{\Delta}}{2}\right)^{2}+M^{2}\right)}\ln\left|\frac{x+\xi}{x-\xi}\,\frac{\left(\vec{k}_{\perp}-\vec{\kappa}_{\perp}\right)^{2}+M^{2}}{\left(\vec{k}-\frac{\vec{\Delta}}{2}\right)^{2}+M^{2}}\right|\,.
\end{eqnarray}

Notice that the sum 
\begin{equation}
H\left(x,\xi,t,\vec{k}_{\perp}\right)=H^{+-}\left(x,\xi,t,\vec{k}_{\perp}\right)+H^{-+}\left(x,\xi,t,\vec{k}_{\perp}\right)\label{eq:A:HTotal}
\end{equation}
is antisymmetric w.r.t. the inversion of the light-cone fraction $x\to-x,$
i.e. $H\left(-x,\xi,t,\vec{k}_{\perp}\right)=-H\left(x,\xi,t,\vec{k}_{\perp}\right).$

We can see that in the points $x=\pm\xi$ the result~(\ref{eq:A:HTotal})
has logarithmic divergences $\sim\ln|x\mp\xi|$. Physically, in this
points one of the quarks has a zero light-cone fraction, and as a
consequence~(\ref{eq:A:HTotal}) becomes very sensitive to the details of short-distance structure of the
model. When we evaluated~(\ref{eq:A:HTotal}), we integrated over
$p^{\pm}$ up to infinity. Rigorously speaking, this contradicts the
basic assumptions of the model, in particular,~(\ref{eq:rho:zerowidth}),
which is valid only when the moments of the active partons are much
smaller than the moment of the whole nucleus. However, since for $p_{1,2}^+\not=0$ the integrals were convergent (the dominant contribution comes from the region where the model is valid), we could ignore such an explicit cutoffs.
Notice that in evaluation of the physical DVCS amplitude~(\ref{CGCF:DVCS:Final}) the cutoffs  $\left|p^-\right|\le q^-/2$ were provided by the external kinematics. Generalization of~(\ref{eq:rho:zerowidth})
to the more realistic color source is a much more complicated task.

\section{$\left\langle U^{\dagger}U\right\rangle $ correlator in finite nuclei.}

\label{sec:A:GluonGPD} As we have seen in the previous section, as
well as we will see in the next section, physical observables depend
on the correlator

\begin{equation}
\left\langle P'\left|U^{\dagger}(x)\, U(y)\right|P\right\rangle \approx\bar{P}^{+}\int d^{3}X\, e^{i\vec{\Delta}\vec{X}}Tr\left(U^{\dagger}(x-X)U(y-X)\right).\end{equation}

Notice that the weight functional $W$ is expressed in terms of the
field $\rho,$ i.e. the correlator is essentially nonlinear object.
In the finite nucleus evaluation of this object slightly differs from
the original derivation given in~\cite{McLerran:1993ni,McLerran:1993ka}.
However, since the weight functional $W[\rho]$ is Gaussian, the total
result can be expressed in terms of the elementary correlator\footnote{To check this, just introduce the external current $J\cdot\rho$ and
evaluate $\langle P'\left|\rho_{1}...\rho_{n}\right|P\rangle$ taking
derivatives. Notice that this would be not true if we had ``interaction
terms'' $\sim\rho^{3},\rho^{4}.$} $\left\langle P'\left|\rho\rho\right|P\right\rangle .$

The final result of our evaluation is\begin{eqnarray}
 &  & S(x,y)=\left\langle P'\left|U^{\dagger}(x_{\perp})\, U(y_{\perp})\right|P\right\rangle =e^{i\vec{\Delta}\frac{\vec{x}_{\perp}+\vec{y}_{\perp}}{2}}\int d^{2}X\, e^{i\vec{\Delta}_{\perp}\vec{X}}\times\label{eq:SDefinition}\\
 &  & \times\exp\left[-g^{2}N_{c}\left(\frac{\tilde{f}\left(\vec{0},\frac{\vec{x}_{\perp}-\vec{y}_{\perp}}{2}-\vec{X}\right)+\tilde{f}\left(\vec{0},-\frac{\vec{x}_{\perp}-\vec{y}_{\perp}}{2}-\vec{X}\right)}{2}-\tilde{f}\left(\vec{x}_{\perp}-\vec{y}_{\perp},-\vec{X}\right)\right)\right],\nonumber \end{eqnarray}

where $\tilde{f}(\vec{r}_{1},\vec{r}_{2})=\int\frac{d^{2}\tilde{\Delta}}{(2\pi)^{2}}e^{-i\tilde{\Delta}\vec{r_{2}}}\int_{-\infty}^{+\infty}dz^{-}\tilde{\gamma}_{A}(z^{-},\vec{r}_{1};\tilde{\Delta}).$

Indeed, using definition

\begin{equation}
U(x)=P\exp\left(ig\int_{-\infty}^{+\infty}dz^{-}\alpha^{a}(z^{-},\vec{x}_{\perp})T_{a}\right),\end{equation}

we may notice that

\begin{itemize}
\item Only the even powers of $\alpha$ give nonzero contribution to (\ref{eq:SDefinition}) 
\item The first term (zero order in $\alpha$) is proportional to $\delta(\Delta)$
and vanishes in the off-forward limit. 
\end{itemize}
Contribution of the second-order term gives

\begin{equation}
-g^{2}N_{c}e^{i\vec{\Delta}_{\perp}\frac{\vec{x}_{\perp}+\vec{y}_{\perp}}{2}}\left(\cos\left(\vec{\Delta}_{\perp}\frac{\vec{x}_{\perp}-\vec{y}_{\perp}}{2}\right)\int dz^{-}\tilde{\gamma}_{A}(z^{-},\vec{0}_{\perp})-\int dz^{-}\tilde{\gamma}_{A}(z^{-},\vec{x}_{\perp}-\vec{y}_{\perp};\Delta)\right).\label{eq:S:2}\end{equation}

It is very convenient to introduce temporary notation $\int dz^{-}\tilde{\gamma}_{A}(z^{-},\vec{r}_{\perp};\Delta)=f(\vec{r}_{\perp};\Delta).$In
this notation (\ref{eq:S:2}) reduces to

\begin{equation}
-g^{2}N_{c}e^{i\vec{\Delta}_{\perp}\frac{\vec{x}_{\perp}+\vec{y}_{\perp}}{2}}\left(\cos\left(\frac{\vec{\Delta}_{\perp}\vec{r}_{\perp}}{2}\right)f(\vec{0}_{\perp};\Delta)-f(\vec{r}_{\perp};\Delta)\right),\end{equation}

where we used notation $\vec{r}=\vec{x}_{\perp}-\vec{y}_{\perp}.$

Evaluation of the higher-order contributions is a bit more tricky.
First we have to notice that the Gaussian form of $W[\rho]$ enables
us to introduce a sort of Wick theorem for evaluation of the multileg
correlators. After that, we have to make Fourier transformation of
each correlator, take the integral over $d^{2}X_{\perp}$ and make
the Fourier back to coordinate space. Performing such procedure step-by-step,
contribution of the $2n$-th order term after some manipulations may
be reduced to

\begin{eqnarray}
 &  & \sum_{m=0}^{2n}\sum_{k=0}^{min(m,2n-m)}\frac{(-1)^{n-m}g^{2n}N_{c}^{n}}{k!(m-k)!(2n-m-k)!}\int\frac{d^{2}\Delta_{1}^{\perp}}{(2\pi)^{2}}...\int\frac{d^{2}\Delta_{n}^{\perp}}{(2\pi)^{2}}\delta\left(\vec{\Delta}_{\perp}-\sum_{i=0}^{n}\vec{\Delta}_{i}^{\perp}\right)\nonumber\\
 &  & \times\left(\prod_{i=1}^{[\frac{m-k}{2}]}f(\vec{0},\vec{\Delta}_{i}^{\perp})e^{i\vec{x}\vec{\Delta}_{i}^{\perp}}\right)\left(\prod_{i=[\frac{m+k}{2}]+1}^{n}f(\vec{0},\vec{\Delta}_{i}^{\perp})e^{i\vec{y}\vec{\Delta}_{i}^{\perp}}\right)\left(\prod_{i=[\frac{m-k}{2}]+1}^{[\frac{m+k}{2}]}f(r_{\perp},\vec{\Delta}_{i}^{\perp})\right)\nonumber \\
 &  &= e^{i\vec{\Delta}\frac{\vec{x}_{\perp}+\vec{y}_{\perp}}{2}}\sum_{m=0}^{2n}\sum_{k=0}^{min(m,2n-m)}\frac{(-1)^{n-m}g^{2n}N_{c}^{n}}{k!(m-k)!(2n-m-k)!}\nonumber \\
 &  &\times \int\frac{d^{2}\Delta_{1}^{\perp}}{(2\pi)^{2}}...\int\frac{d^{2}\Delta_{n}^{\perp}}{(2\pi)^{2}}\delta(\vec{\Delta}_{\perp}-\sum_{i=0}^{n}\vec{\Delta}_{i}^{\perp})\nonumber \\
 &  & \times\left(\prod_{i=1}^{[\frac{m-k}{2}]}f(\vec{0},\vec{\Delta}_{i}^{\perp})e^{i\vec{r}\vec{\Delta}_{i}^{\perp}/2}\right)\left(\prod_{i=[\frac{m+k}{2}]+1}^{n}f(\vec{0},\vec{\Delta}_{i}^{\perp})e^{-i\vec{r}\vec{\Delta}_{i}^{\perp}/2}\right)\left(\prod_{i=[\frac{m-k}{2}]+1}^{[\frac{m+k}{2}]}f(r_{\perp},\vec{\Delta}_{i}^{\perp})\right).\label{eq:S:temp:multi} \end{eqnarray}

Now we replace back $\delta(\vec{\Delta}_{\perp}-\sum_{i=0}^{n}\vec{\Delta}_{i}^{\perp})=\int d^{2}X\, e^{i\vec{\Delta}_{\perp}\vec{X}_{\perp}}\prod_{i=1}^{n}e^{-i\vec{\Delta}_{i}^{\perp}\vec{X}_{\perp}}$ and
reduce (\ref{eq:S:temp:multi}) to

\begin{eqnarray}
 &  & e^{i\vec{\Delta}\frac{\vec{x}_{\perp}+\vec{y}_{\perp}}{2}}\int d^{2}X\, e^{i\vec{\Delta}_{\perp}\vec{X}_{\perp}}\sum_{m=0}^{2n}\sum_{k=0}^{min(m,2n-m)}\frac{(-1)^{n-m}g^{2n}N_{c}^{n}}{k!(m-k)!(2n-m-k)!}\nonumber\\
 &  &\nonumber \times\left(\prod_{i=1}^{[\frac{m-k}{2}]}\int\frac{d^{2}\Delta_{i}^{\perp}}{(2\pi)^{2}}f(\vec{0},\vec{\Delta}_{i}^{\perp})e^{i\vec{r}\vec{\Delta}_{i}^{\perp}/2}\right)\left(\prod_{i=[\frac{m+k}{2}]+1}^{n}\int\frac{d^{2}\Delta_{i}^{\perp}}{(2\pi)^{2}}f(\vec{0},\vec{\Delta}_{i}^{\perp})e^{-i\vec{r}\vec{\Delta}_{i}^{\perp}/2}\right)\\
 &  &\nonumber
\times\left(\prod_{i=[\frac{m-k}{2}]+1}^{[\frac{m+k}{2}]}\int\frac{d^{2}\Delta_{i}^{\perp}}{(2\pi)^{2}}f(r_{\perp},\vec{\Delta}_{i}^{\perp})\right)\\
 &  & \nonumber
=e^{i\vec{\Delta}\frac{\vec{x}_{\perp}+\vec{y}_{\perp}}{2}}\int d^{2}X\, e^{i\vec{\Delta}_{\perp}\vec{X}}\sum_{m=0}^{2n}\sum_{k=0}^{min(m,2n-m)}\frac{(-1)^{n-m}g^{2n}N_{c}^{n}}{k!(m-k)!(2n-m-k)!}\\
 &  &\nonumber
\times\tilde{f}^{[\frac{m-k}{2}]}\left(\vec{0},\frac{\vec{r}}{2}-\vec{X}\right)\tilde{f}^{[\frac{m+k}{2}]}\left(\vec{0},-\frac{\vec{r}}{2}-\vec{X}\right)\tilde{f}^{k}\left(\vec{0},-\vec{X}\right)\\
&  &
=e^{i\vec{\Delta}\frac{\vec{x}_{\perp}+\vec{y}_{\perp}}{2}}\int d^{2}X\, e^{i\vec{\Delta}_{\perp}\vec{X}}\exp\left[-g^{2}N_{c}\left(\frac{\tilde{f}\left(\vec{0},\frac{\vec{r}}{2}-\vec{X}\right)+\tilde{f}\left(\vec{0},-\frac{\vec{r}}{2}-\vec{X}\right)}{2}-\tilde{f}\left(\vec{r},-\vec{X}\right)\right)\right],
\end{eqnarray}

in agreement with (\ref{eq:SDefinition}).

For evaluation of the complicated objects like $\left\langle P'\left|\Phi[\rho]U^{\dagger}(x_{\perp})\, U(y_{\perp})\right|P\right\rangle $
(see e.g. Gluon distributions) we can use a quasiclassical
formula

\begin{eqnarray}
&  & \left\langle P'\left|\hat{A}(x,y)\hat{B}(x,y)\right|P\right\rangle =\frac{e^{i\vec{\Delta}\frac{\vec{x}+\vec{y}}{2}}}{\bar{P}^{+}}\int d^{3}X\, e^{i\vec{\Delta}\vec{X}}\nonumber\\
&  & \times\left(\int\frac{d^{3}\Delta_{1}}{(2\pi)^{3}}\, e^{-i\vec{\Delta}_{1}\vec{X}}\left\langle P+\Delta_{1}\left|\hat{A}\left(\frac{\vec{r}}{2},-\frac{\vec{r}}{2}\right)\right|P\right\rangle \right)\left(\int\frac{d^{3}\Delta_{2}}{(2\pi)^{3}}\, e^{-i\vec{\Delta}_{2}\vec{X}}\left\langle P+\Delta_{2}\left|\hat{B}\left(\frac{\vec{r}}{2},-\frac{\vec{r}}{2}\right)\right|P\right\rangle \right)\nonumber\\
&  &= \frac{e^{i\vec{\Delta}\frac{\vec{x}+\vec{y}}{2}}}{\bar{P}^{+}}\int\frac{d^{3}\Delta_{1}}{(2\pi)^{3}}\frac{d^{3}\Delta_{2}}{(2\pi)^{3}}\,(2\pi)^{3}\delta^{3}(\Delta-\Delta_{1}-\Delta_{2})\left\langle P+\Delta_{1}\left|\hat{A}\left(\frac{\vec{r}}{2},-\frac{\vec{r}}{2}\right)\right|P\right\rangle \left\langle P+\Delta_{2}\left|\hat{B}\left(\frac{\vec{r}}{2},-\frac{\vec{r}}{2}\right)\right|P\right\rangle\nonumber \\
&  & =e^{i\vec{\Delta}\frac{\vec{x}+\vec{y}}{2}}\int d^{3}X\, e^{i\vec{\Delta}\vec{X}}A^{cl}\left(\frac{\vec{r}}{2}-\vec{X},-\frac{\vec{r}}{2}-\vec{X}\right)B^{cl}\left(\frac{\vec{r}}{2}-\vec{X},-\frac{\vec{r}}{2}-\vec{X}\right),
\end{eqnarray}

where $\vec{r}=\vec{x}-\vec{y.}$

 \end{document}